\documentclass[12pt]{article} 
\usepackage{amstex}
\usepackage{amssymb}
\usepackage{citesort}
\usepackage{graphicx}                              
\usepackage[nomarkers]{endfloat}
\usepackage{pstricks}
\renewcommand{\theequation}{\thesection.\arabic{equation}}
\begin{document}

%\begin{titlepage}

\vspace*{2.5cm}

\begin{center}
{\Large
Dual superconductivity in the SU(2) pure gauge vacuum:\\[0.3cm]
a lattice study}
\end{center}

\vspace*{2cm}

\renewcommand{\thefootnote}{\fnsymbol{footnote}}

\begin{center}
{\large
Paolo Cea$^{1,2,}$\protect\footnote{Electronic address:
cea@@bari.infn.it} and
Leonardo Cosmai$^{2,}$\protect\footnote{Electronic address:
cosmai@@bari.infn.it} \\[0.5cm]
$^1$Dipartimento di Fisica dell'Universit\`a di Bari,
70126 Bari, Italy\\[0.3cm]
$^2$Istituto Nazionale di Fisica Nucleare, Sezione di Bari,
70126 Bari, Italy
}
\end{center}

\vspace*{0.5cm}

\begin{center}
{\large
April 10, 1995
}
\end{center}

\vspace*{1.0cm}

\renewcommand{\abstractname}{\normalsize Abstract}
\begin{abstract}
We investigate the dual superconductivity hypothesis in pure SU(2)
lattice gauge theory. We focus on the dual Meissner effect by
analyzing the distribution of the color fields due to a static
quark-antiquark pair. We find evidence of the dual Meissner effect
both in the maximally Abelian gauge and without gauge fixing. We
measure the London penetration length. Our results suggest that the
London penetration length is a physical gauge-invariant quantity. We
put out a simple relation between the penetration length and the
square root of the string tension. We find that our extimation is
quite close to the extrapolated continuum limit available in the
literature. A remarkable consequence of our study is that an effective
Abelian theory can account for the long range properties of the SU(2)
confining vacuum.   
\end{abstract}
\vspace*{0.5cm}
hep-lat/9504008

{
\thispagestyle{myheadings}
\renewcommand{\thepage}{BARI - TH 195/95}
\clearpage
}
\renewcommand{\thesection}{\Roman{section}.}
\section{INTRODUCTION}
\label{Section_I}
\renewcommand{\thesection}{\arabic{section}}

Understanding the mechanism of quark confinement is a central problem
in the high energy physics. This requires, among other things, to
identify the dynamical variables which are relevant to the
confinement.

A satisfying solution would be to set up an approximate vacuum state
which confines color charges. This way one could derive an effective
action which describe the long-distance properties of
QCD~\cite{Baker91}. Even this incomplete program, however, mandates a
non-perturbative approach. Fortunately we have at our disposal a
framework in which we can do non-perturbative calculations, namely the
lattice discretization of gauge theories. Since a typical Monte Carlo
simulation generates vacuum configurations, one expects to gain
information on the non-perturbative vacuum structure.

However a guideless search into the numerical configurations generated
during Monte Carlo runs is hopeless. In other words, we need some
theoretical input which selects the dynamical variables relevant to
the confinement. The situation looks similar to the theory of
superconductivity. Indeed, it was the Cooper's observation that the
Fermi surface is unstable with regard to the formation of bounded
electron pairs which led  Bardeen, Cooper, and Schrieffer to formulate
the successful BCS superconductivity theory~\cite{Schrieffer92}.

An interesting possibility has been conjectured long time ago by
G.~'t~Hooft~\cite{tHooft75} and S.~Mandelstam ~\cite{Mandelstam76}.
These authors proposed that the confining vacuum behaves as a coherent
state of color magnetic monopoles. This is equivalent to say that the
vacuum is a magnetic (dual) superconductor. This fascinating proposal
offers a picture of confinement whose physics can be clearly
extracted. As a matter of fact, the dual Meissner effect causes the
formation of chromoelectric flux tubes between chromoelectric charges
leading to a linear rising potential. It is worthwhile to discuss
briefly the 't~Hooft's proposal~\cite{Nota4}.

Let us consider the non-Abelian gauge theory spontaneously broken via
the Higgs mechanism. The Higgs fields are in the adjoint
representation. For concreteness we focus on the Georgi-Glashow
model~\cite{Georgi72}. It is well known that the Georgi-Glashow model
allows field configurations which correspond to magnetic
monopoles~\cite{Monopoles}. Moreover, one readily finds that the
monopole mass is  given by
\begin{equation}
\label{monopole_mass}
M_{\text{mon}} = C \, \frac{M_W}{\alpha}   \;,
\end{equation}
where $M_W$ is the mass of the charged vector boson, $C$ a constant
and $\alpha$ the fine structure constant. The dual superconductor
scenario is realized if these magnetic monopoles condense by means of
the magnetic Higgs mechanism. This means that the monopoles become
tachionic:
\begin{equation}
\label{tachionic_monopole}
M^2_{\text{mon}} \le 0 \qquad .
\end{equation}
From Equation\eqref{monopole_mass} we see that $M^2_W \rightarrow 0$
(if we kept $\alpha$ fixed). The fact that $M^2_W$ must go through
zero suggests that the original Higgs field could be removed. Thus we
are led to consider the pure gauge theory without elementary Higgs
fields. The role of the scalar Higgs field is played by any operator
which transforms in the adjoint representation of the gauge group.
More precisely, after choosing an operator $X(x)$ which transforms
according to the adjoint representation, one fixes the gauge by
diagonalizing $X(x)$ at each point. This choice does not fix the gauge
completely; it leaves as residual invariance group the maximally
Abelian (Cartan) subgroup of the gauge group. Such a procedure is
known as Abelian projection~\cite{tHooft81}. For instance, if the gauge
group is SU(N), then after gauge fixing the residual invariance group
is $\text{U(1)}^{N-1}$. The world line of the monopoles can be
identified as the lines where two eigenvalues of the operator $X(x)$
are equal. The dual superconductor idea is realized if these Abelian
monopoles condense. 

It is evident that the monopoles are dynamical; they will take part
in the dynamics of the system. As a consequence the problem of
monopole condensation cannot be dealt with the perturbation theory. On
the other hand, the Abelian projection can be implemented on the
lattice~\cite{Kronfeld87} . Thus one can analyze the dynamics of the
Abelian projected gauge fields by means of Monte Carlo simulations. In
the following we shall consider the pure SU(2) gauge theory.

To perform the Abelian projection we make a choice for $X(x)$. The
simpler possibility is to consider a local quantity. For instance, we
can use a plaquette with a definite orientation (field-strength gauge)
or the Polyakov loop (Polyakov gauge). In these unitary gauges we
implement the gauge fixing by means of the matrices $V(x)$ which
diagonalize $X(x)$ at each lattice site:
\begin{equation}
\label{X(x)}
V(x) X(x) V^\dagger(x) = \text{diag} \left[ e^{i \alpha(x)}, e^{-i
\alpha(x)} \right] \quad .
\end{equation}
It is straightforward to check that the residual gauge invariance
group is the U(1) group with transformations $\exp(i \sigma_3
\theta(x))$.

The Abelian projection of the gauge transformed links 
\begin{equation}
\label{Utilde}
\widetilde{U}_\mu(x) = V(x) U_\mu(x) V^\dagger(x+\hat{\mu})
\end{equation}
amounts to write
\begin{equation}
\label{WaU}
\widetilde{U}_\mu(x) = W_\mu(x) U^A_\mu(x)
\end{equation}
with
\begin{equation}
\label{Ua}
U^A_\mu(x) =  \text{diag} \left[ e^{i \theta^A_\mu(x)}, 
e^{-i \theta^A_\mu(x)} \right] \quad ,
\end{equation}
\begin{equation}
\label{theta}
\theta^A_\mu(x) =  \text{arg} \left[ \widetilde{U}_\mu(x) 
\right]_{11} \quad .
\end{equation}
$U^A_\mu(x)$ is the Abelian projection of $\widetilde{U}_\mu(x)$.

A different class of gauge fixing has been proposed in the
literature, namely the Abelian covariant gauge or maximally Abelian
gauge~\cite{Kronfeld87}. In the continuum the maximally Abelian gauge
corresponds to impose the constraints~\cite{vanderSijs91}:
\begin{equation}
\label{DmuAmu}
D_\mu A^\pm_\mu(x) = 0
\end{equation}
where $A^\pm_\mu = A^1_\mu \pm i A^2_\mu$, and $D_\mu$ is the
$A^3_\mu$-covariant derivative. On the lattice the
constraints~(\ref{DmuAmu}) can be implemented like the Landau
gauge~\cite{Mandula90}. Indeed Equation (\ref{DmuAmu}) corresponds on
the lattice to diagonalize~\cite{vanderSijs91,Suzuki90}
\begin{equation}
\label{diagonalization}
X(x) = \sum_\mu \left\{ U_\mu(x) \sigma_3 U^\dagger_\mu(x) +
U^\dagger_\mu(x-\hat{\mu}) \sigma_3   U_\mu(x-\hat{\mu}) \right\} \,
.
\end{equation}
To do this it is enough to maximize iteratively the quantity 
\begin{equation}
\label{R}
R = \sum_{x, \mu} \left[ \sigma_3 \widetilde{U}_\mu(x) \sigma_3
\widetilde{U}^\dagger_\mu(x) \right] \; ,
\end{equation}
where the $\widetilde{U}_\mu(x)$'s are the gauge transformed
links~(\ref{Utilde}). We thereby obtain the matrices $V(x)$ and
perform the Abelian projection of the links by
Eqs.~(\ref{Utilde})-(\ref{theta}).

From the above discussion it is evident that the monopole dynamic does
depend on the choice of the operator needed to fix the gauge. On the
other hand the confinement of color charges via monopole condensation
can not depend on the gauge fixing. However, it is conceivable that
the dual superconductor scenario could manifest with a judicious
choice of $X(x)$. This outcome could arise from a gauge fixing which
freezes the degrees of freedom which are irrelevant to the
confinement. We feel that the situation is similar to the time-honored
BCS theory of superconductivity. Indeed in the BCS theory one deals
with a reduced Hamiltonian which breaks the electromagnetic gauge
invariance. Nevertheless, the reduced BCS Hamiltonian offered the
correct explanation of the Meissner effect. As a matter of fact, it
was showed~\cite{Anderson58} that the collective states which are
essential to restore the gauge invariance do not contribute to the BCS
calculation of the Meissner effect. In other words, the reduced BCS
Hamiltonian, by retaining the degrees of freedom relevant to the
superconductivity, gives a sensible answer even though it breaks the
electromagnetic gauge invariance.

Interesting enough, it turns out that, if one fixes the maximally
Abelian gauge, the Abelian projected links seem to retain the
informations relevant to the confinement~\cite{Suzuki93}. Thus, it is
important to deepen the study of the dynamics of the Abelian projected
fields in that particular gauge fixing.

The aim of the present paper is to analyze the finger-print of the
dual superconductor hypothesis, namely the Meissner effect. To this
end, we analyze the distribution of the color field due to static
quark-antiquark pair in SU(2) lattice gauge theory in the maximally
Abelian gauge. Moreover we will study the gauge dependence of the
London penetration length. A partial account of this paper has been
published in Ref.~\cite{Cea95}.

The plan of the paper is as follows. In Sec. II we explore the field
configurations produced by the quark-antiquark static pairs both in
the case of Abelian projected links after maximally Abelian gauge has
been fixed, and in the case of full SU(2) links. In Sec.~III we
analyze the transverse distribution of the longitudinal chromoelectric
field. In Sec. IV we investigate the relation between the penetration
length and the string tension. Our conclusions are relegated in Sec.
V. The Appendix comprises several technical details on the maximally
Abelian gauge fixing.

\renewcommand{\thesection}{\Roman{section}.}
\setcounter{equation}{0}
\section{COLOR FIELDS}
\label{Section_II}
\renewcommand{\thesection}{\arabic{section}}

In this Section we analyze the distribution of the color fields due to
static quark-antiquark pairs. Following the authors of
Ref.~\cite{DiGiacomo90}, we can measure the color fields by means of
the correlation of a plaquette $U_p$ with a Wilson loop $W$. The
plaquette is connected to the Wilson loop by a Schwinger line L (see
Fig.~\ref{Fig:correlator}). Moving the plaquette $U_p$ with respect to
the Wilson loop one can scan the structure of the color fields. In a
previous study~\cite{Cea93} we found evidence of the dual Meissner
effect in the maximally Abelian gauge. In particular we measured the
penetration depth of the flux tube chromoelectric field. However in
Ref.~\cite{Cea93} we employed rather small lattices ($L=12$). In this
work we extend our previous study in two directions. Firstly, we
perform numerical simulations on lattice whose size ranges from $L=16$
up to $L=24$. In addition we investigate the gauge invariance of the
penetration length. To do this we perform the numerical simulations
both in the maximally Abelian gauge and without gauge fixing.

\renewcommand{\thesubsection}{\Alph{subsection}.}
\subsection{SU(2)}
\label{Section_II_A}

According to Ref.~\cite{DiGiacomo90}, one can explore the field
configurations produced by the quark-antiquark pair by measuring the
connected correlation function (Fig.~\ref{Fig:correlator})
\begin{equation}
\label{rhoW}
\rho_W = \frac{\left\langle \text{tr}
\left( W L U_P L^{\dagger} \right)  \right\rangle}
              { \left\langle \text{tr} (W) \right\rangle }
 - \frac{1}{2} \,
\frac{\left\langle \text{tr} (U_P) \text{tr} (W)  \right\rangle}
              { \left\langle \text{tr} (W) \right\rangle } \; ,
\end{equation}
where $U_P=U_{\mu\nu}(x)$ is the plaquette in the $(\mu,\nu)$ plane.
Note that the correlation function~(\ref{rhoW}) is sensitive to the
field strength rather than to the square of the field
strength~\cite{DiGiacomo94}:
\begin{equation}
\label{rhoWlimcont}
\rho_W  @>>{a \rightarrow 0}>a^2 g \left[ \left\langle
F_{\mu\nu}\right\rangle_{q\bar{q}} - \left\langle F_{\mu\nu}
\right\rangle_0 \right]  \;.
\end{equation}
According to Eq.(\ref{rhoWlimcont}) we define the color field strength
tensor as:
\begin{equation}
\label{fieldstrength}
F_{\mu\nu}(x) = \frac{\sqrt{\beta}}{2} \, \rho_W(x)   \;.
\end{equation}
By varying the distance and the orientation of the plaquette $U_P$
with respect to the Wilson loop $W$, one can scan the color field
distribution of the flux tube.

We performed numerical simulations  with Wilson action and periodic
boundary conditions using an overrelaxed Metropolis algorithm. Our
data refer to $16^4$, $20^4$, and  $24^4$ lattices. To evaluate the
correlator Eq.~(\ref{rhoW}) we used square Wilson loop $L_W \times
L_W$, with $L_W=L/2 -2$ ($L$ being the lattice size), and rectangular
Wilson loops $L/2 \times L/4$.

In order to reduce the quantum fluctuations we adopted the controlled
cooling algorithm~\cite{Campostrini90}. It is
known~\cite{Campostrini89} that by cooling in a smooth way equilibrium
configurations, quantum fluctuations are reduced by a few order of
magnitude, while the string tension survives and shows a plateau. We
shall show below that the penetration length behaves in a similar way.

For reader convenience let us, briefly, illustrate our cooling
procedure. The lattice  gauge configurations are cooled by replacing the
matrix $U_\mu(x)$ associated to each link $l\equiv(x,\hat{\mu})$ with
a new matrix $U_\mu^\prime(x)$ in such a way that the local
contribution to the lattice action
\begin{equation}
\label{DeltaS}
S(x) = 1 - \frac{1}{2} \text{tr} \left\{ U_\mu(x) k(x) F(x)
\right\}
\end{equation}
is minimized. $\widetilde{F}(x) = k(x) F(x)$ is the sum over the
``U-staples'' involving the link $l$ and
$k(x)=\sqrt{\text{det}\left(\widetilde{F}(x)\right)}$, so that
$F(x)\in$~SU(2). In a ``controlled'' or ``smooth'' cooling step we
have
\begin{equation}
\label{controlled_cooling}
U_\mu(x) \rightarrow  U^\prime_\mu(x) = V(x) U_\mu(x) \;,
\end{equation}
where $V(x)$ is the SU(2) matrix which maximizes 
\begin{equation}
\label{Vmax}
\text{tr} \left\{ V(x) U_\mu(x) F(x) \right\}
\end{equation}
subjected to the following constraint on the SU(2) distance between
$U_\mu(x)$ and $U^\prime_\mu(x)$:
\begin{equation}
\label{constraint}
\frac{1}{4} \text{tr} \left[ \left( U_\mu^\dagger(x) -
U^{\prime\dagger}_\mu(x) \right)  \left( U_\mu(x) - U^\prime_\mu(x)
\right) \right] \; \le \delta^2 \,.
\end{equation}
We adopt $\delta=0.0354$. A complete cooling sweep consists in the
replacement Eq.~\eqref{controlled_cooling} at each lattice site. We do
the above replacement vector-like according to the standard
checkerboard order.

The cooling technique allows us to disentangle the signal from the
noise with a relatively small statistics. After discarding about 3000 
sweeps to insure thermalization we collect measurements on
configurations separated by 100 upgrades for 9 different values of
$\beta$ in the range  $2.45 \le \beta \le 2.7$ After cooling we
obtained a good signal for $\rho_W$  on very small statistical samples
($20 \div 100$ configurations).

\begin{figure}
\includegraphics[width=\textwidth]
{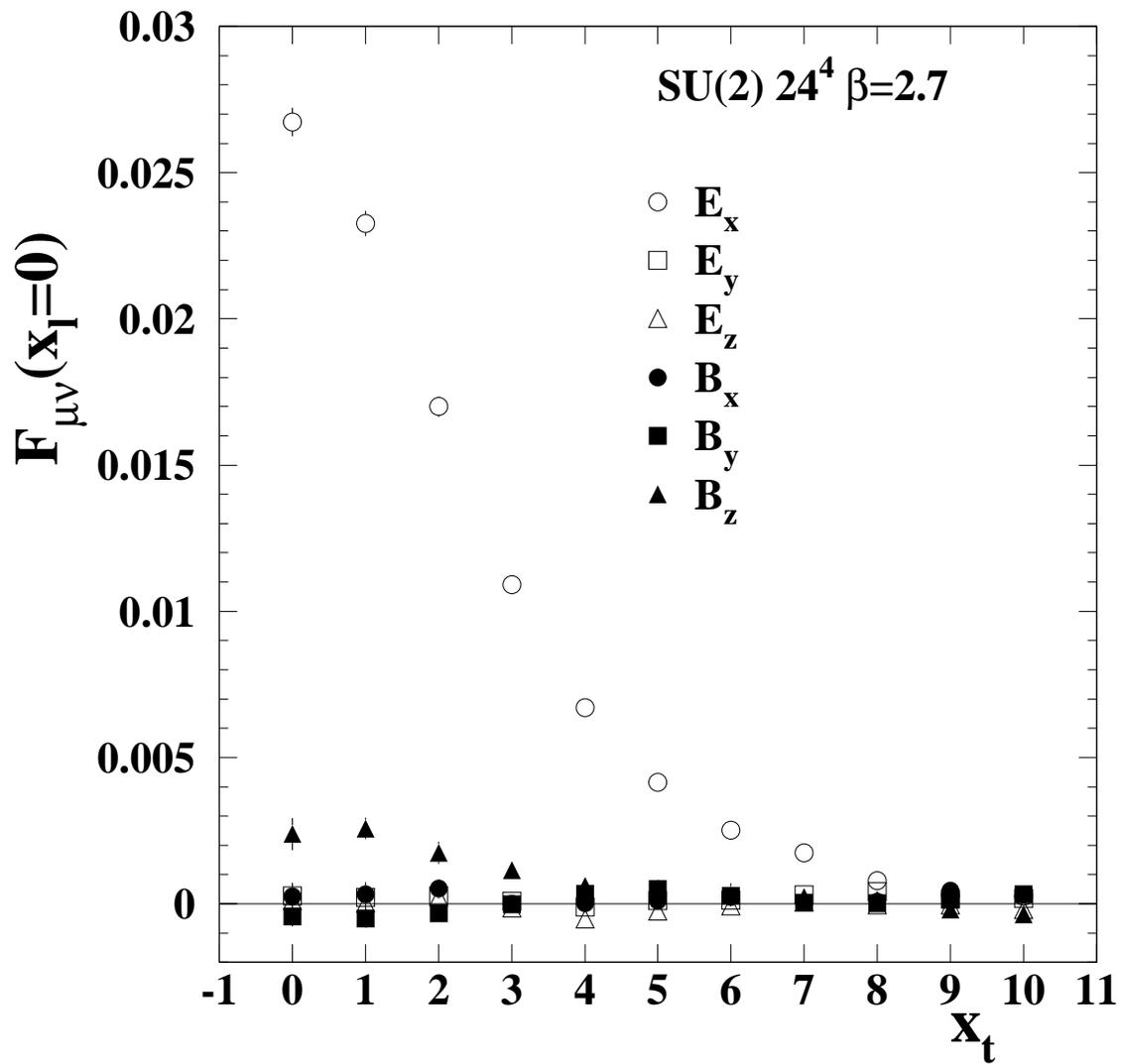}
\caption{The field strength tensor $F_{\mu\nu}(x_l,x_t)$ evaluated at
$x_l=0$ on a $24^4$ lattice at $\beta=2.7$, using Wilson loops of size
$10 \times 10$ in Eq.~\protect\eqref{rhoW}.}
\label{Fig:Fmunu}
\end{figure}

In Figure~\ref{Fig:Fmunu} we report our results for the field strength
tensor $F_{\mu\nu}(x_l,x_t)$, where the coordinates $x_l$, $x_t$
measure respectively the distance from the middle point between quark
and antiquark (which corresponds to the center of the spatial side of
the Wilson loop $W$ in Eq.~(\ref{rhoW})) and the distance out of the
plane defined by the Wilson loop.

\begin{figure}
\includegraphics[width=\textwidth]
{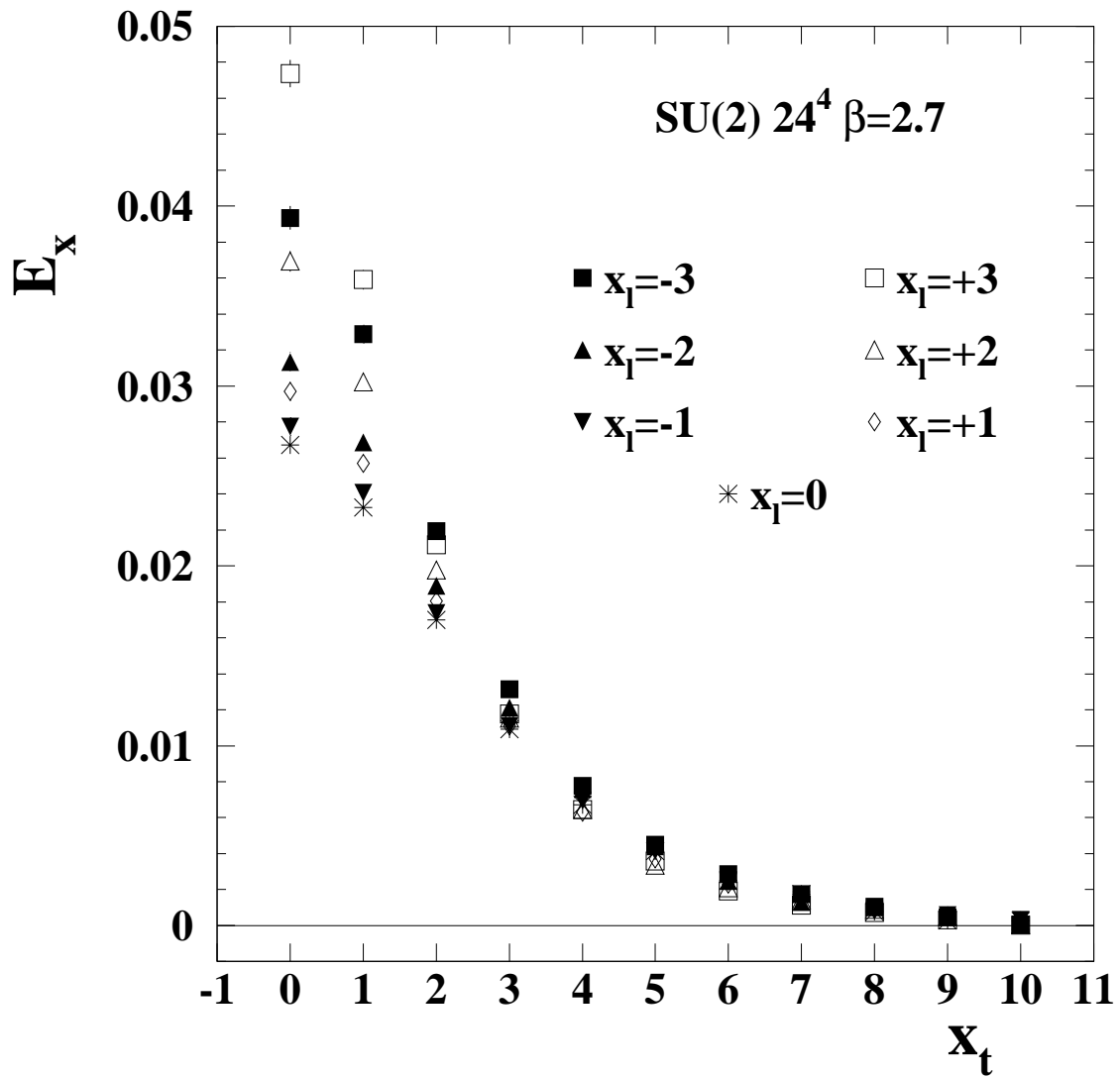}
\caption{The $x_l$-dependence of the transverse profile of the
longitudinal chromoelectric field $E_x(x_l,x_t) \equiv E_l(x_l,x_t)$.} 
\label{Fig:El(xt)}
\end{figure}

The entries  in Fig.~\ref{Fig:Fmunu} refer
to measurements of the field strength tensor taken in the middle of
the flux tube ($x_l=0$) with 8 cooling steps at $\beta=2.7$ on the
$24^4$ lattice, using a square Wilson loop $W$ of size $10 \times 10$.
Our results show that $\rho_W$ is sizeable when $U_p$ and $W$ are in
parallel planes. This corresponds to measure the component $E_l$ of
the chromoelectric field directed along the line joining the
$q\bar{q}$ pair ($E_x$ in Fig.~\ref{Fig:Fmunu}). Moreover we see that 
$E_l(x_l,x_t)$ decreases rapidly in the transverse direction $x_t$. 
In Figure~\ref{Fig:El(xt)} we display the transverse distribution of
the longitudinal chromoelectric field along the flux tube. The static
color sources are at $x_l=+5$ and $x_l=-4$ (in lattice units).
Figure~\ref{Fig:El(xt)} shows that the effects of the color sources on
the chromoelectric fields extends over about three lattice spacings.
Remarkably, far from the sources the longitudinal chromoelectric field
is almost constant along the $q - \overline{q}$ line. Thus, the color
field structure of the $q-\bar{q}$ tube which emerges from our results
is quite simple: the flux tube is almost completely formed by the
longitudinal chromoelectric field which is constant along the flux
tube (if $x_l$ is not too close to the static color sources) and
decreases rapidly in the transverse direction.

\renewcommand{\thesubsection}{\Alph{subsection}.}
\subsection{Maximally Abelian projection}
\label{Section_II_B}

\begin{figure}
\includegraphics[width=\textwidth]
{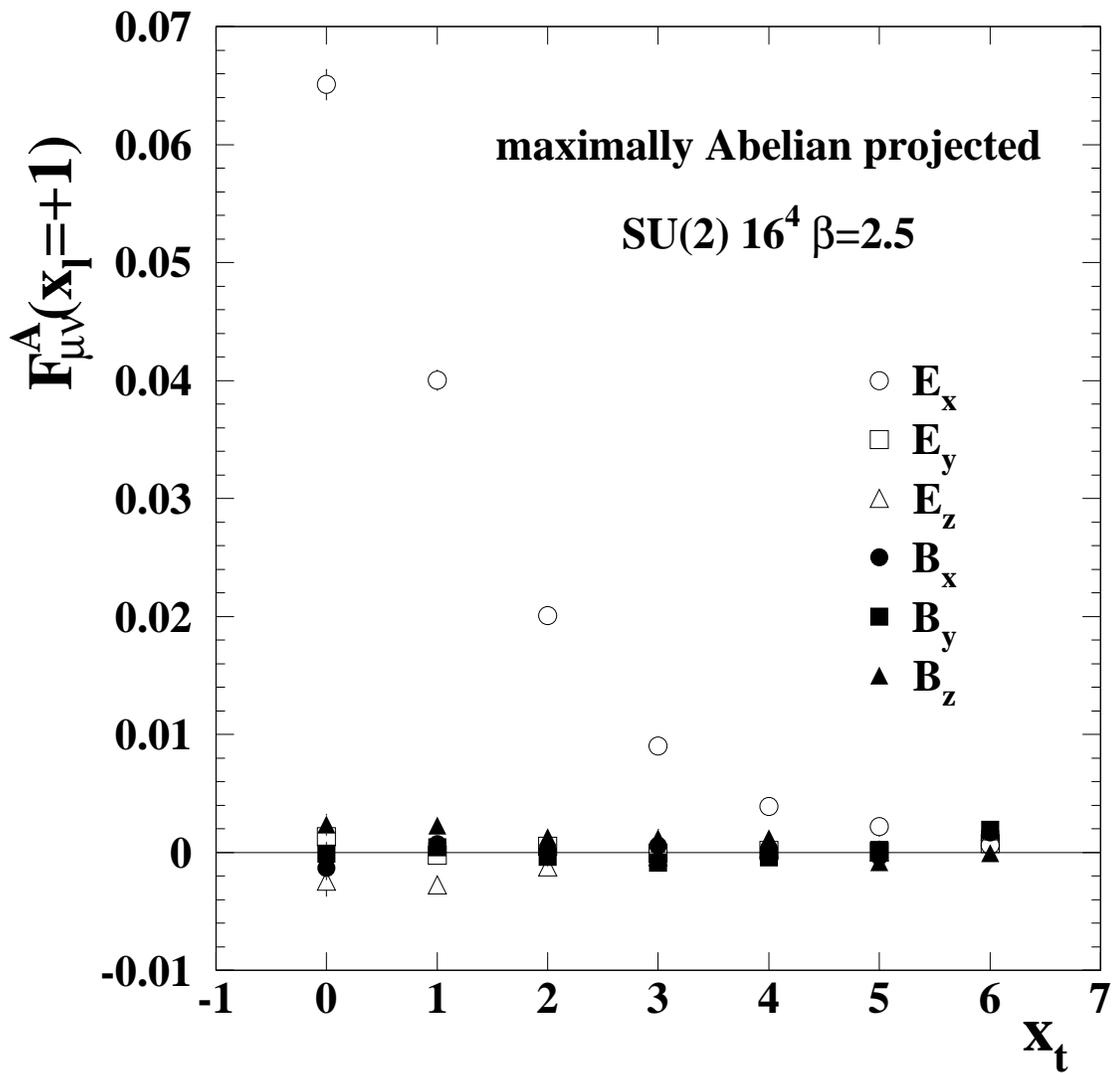}
\caption{The maximally Abelian projected field strength tensor
$F^A_{\mu\nu}(x_l,x_t)$ evaluated at $x_l=+1$ on a $16^4$ lattice at
$\beta=2.5$, using Wilson loops of size $6 \times 6$ in
Eq.~\protect\eqref{rhoWab}.}
\label{Fig:FmunuAb}
\end{figure}

\begin{figure}
\includegraphics[width=\textwidth]
{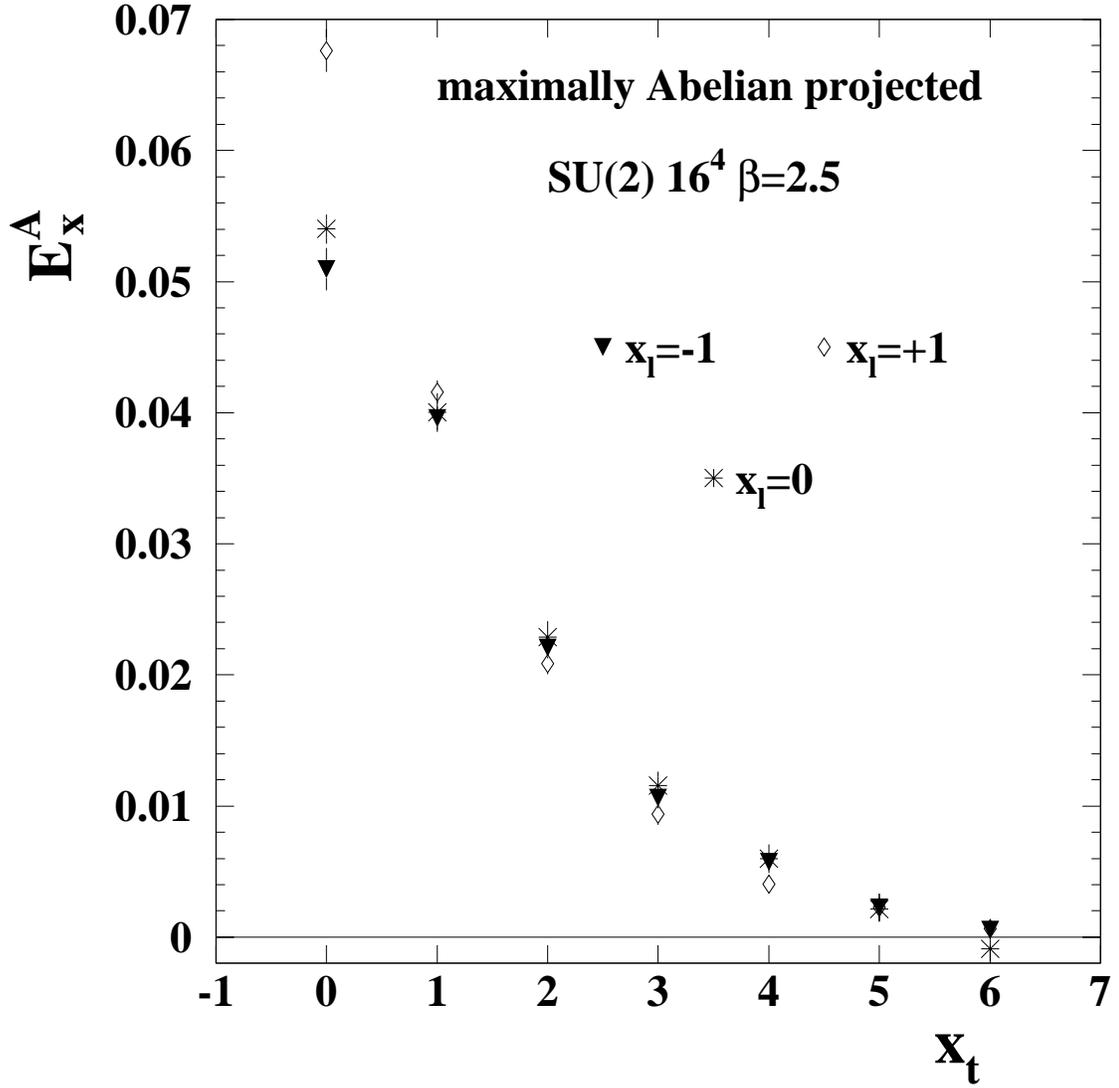}
\caption{The maximally Abelian projected longitudinal chromoelectric
field $E^A_x(x_l,x_t) \equiv E^A_l(x_l,x_t)$ versus the transverse
distance from the flux tube $x_t$ for three different values of the
longitudinal coordinate.}
\label{Fig:El_Ab(xt)}
\end{figure}

In the 't~Hooft formulation~\cite{tHooft81} the dual superconductor
model is elaborated through the Abelian projection. The idea is
that the Abelian projected gauge fields retain the long distance
physics of the gauge system. In particular the physical quantities
related to the confinement should be independent of the gauge fixing,
and agree with those obtained in the full gauge system. This suggested
us~\cite{Cea93} to investigate the Abelian projected correlator
\begin{equation}
\label{rhoWab}
\rho_W^{A} = \frac{\left\langle \text{tr}
\left(  W^A  U_P^A \right) \right\rangle}
              { \left\langle \text{tr} \left( W^A  \right)
\right\rangle }
 - \frac{1}{2} \,
\frac{\left\langle \text{tr}
\left( U_P^A \right) \text{tr} \left( W^A  \right) \right\rangle}
              { \left\langle \text{tr}
\left( W^A \right) \right\rangle } \; .
\end{equation}
The correlator $\rho_W^A$ is obtained from Eq.\eqref{rhoW} with the
substitution $U_\mu(x) \rightarrow U^A_\mu(x)$. For instance the
Abelian projected plaquette in the $(\mu,\nu)$ plane is 
\begin{equation}
\label{Abplaq}
\begin{split}
U^A_{\mu\nu}(x) &=  U^A_\mu(x) U^A_\nu(x+\hat{\mu})
U^{A\dagger}_\mu(x+\hat{\nu}) U^{A\dagger}_\mu(x) \\
&= \text{diag}\left[\exp\left(i \theta^A_{\mu\nu}(x)\right), 
\exp\left(-i \theta^A_{\mu\nu}(x)\right) \right] \;.
\end{split}
\end{equation}
Obviously the Abelian projected quantities are commutating, so that we
do not need the Schwinger lines in Eq.\eqref{rhoWab}. It is worthwhile
to stress that $\rho^A_W$ is a gauge-dependent correlator. We
performed measurement for 6 different values of $\beta$ in the range
$2.45 \le \beta \le 2.70$ using the $16^4$ and $20^4$ lattices. In
this case we find a good signal without cooling. Measurements are
taken on a sample of $500-700$ configurations each separated by $50$
upgrades, after discarding 3000 sweeps to allow thermalization. The
maximally Abelian gauge is fixed iteratively via the overrelaxation
algorithm of Ref.~\cite{Mandula90} with the overrelaxation parameter
$\omega=1.7$ (for further details see the Appendix). Remarkably
enough, it turns out that the Abelian field strength tensor 
\begin{equation}
\label{FmunuAb}
F^A_{\mu\nu}(x) = \frac{\sqrt{\beta}}{2} \rho^A_W(x)
\end{equation}
behaves like the gauge-invariant one defined by
Eq.\eqref{fieldstrength}. In Figure~\ref{Fig:FmunuAb} we report our
results for the field strength tensor $F_{\mu\nu}(x_l,x_t)$ evaluated
on maximally Abelian projected gauge configurations. The entries in
Figure~\ref{Fig:FmunuAb} refer to measurements done at $x_l=+1$ on a
$16^4$ lattice at $\beta=2.5$ using a square Wilson loop of size $6
\times 6$ in Eq.~\eqref{rhoWab}. Again we see that only the
longitudinal chromoelectric field is sizeable. In
Figure~\ref{Fig:El_Ab(xt)} we study the $x_l$-dependence of the
longitudinal Abelian  chromoelectric field  extracted using $6 \times
6$ Wilson loop in Eq.\eqref{rhoW} at $\beta=2.5$ on the $16^4$
lattice. Note that in the present case the static sources are at
$x_l=+3$ and $x_l=-2$. The longitudinal Abelian chromoelectric field, 
likewise the non Abelian one, does not depend on the longitudinal
coordinate $x_l$, far from the static sources. It is worthwhile to
observe that Fig.~\ref{Fig:El_Ab(xt)} suggests that the Abelian static
sources are more localized than the non Abelian ones. This is in
accordance with our previous observation that the maximally Abelian
gauge fixing seems to reduce the fluctuations which are unimportant
for the long distance physics. 

In the next Section we shall analyze our numerical data within the
dual superconductor hypothesis.

\renewcommand{\thesection}{\Roman{section}.}
\setcounter{equation}{0}
\section{LONDON PENETRATION LENGTH}
\label{Section_III}
\renewcommand{\thesection}{\arabic{section}}

\renewcommand{\thesubsection}{\Alph{subsection}.}
\subsection{SU(2)}
\label{Section_III_A}

If the dual superconductor scenario holds, the transverse shape of the
longitudinal chromoelectric field $E_l$ should resemble the dual
version of the Abrikosov vortex field distribution. Hence we expect
that $E_l(x_t)$ can be fitted according to 
\begin{equation}
\label{London}
E_l(x_t) = \frac{\Phi}{2 \pi} \mu^2 K_0(\mu x_t) \;, x_t > 0
\end{equation}
where $K_0$ is the modified Bessel function of order zero, $\Phi$ is
the external flux, and $\lambda=1/\mu$ is the London penetration
length. Equation~\eqref{London} is valid if $\lambda \gg \xi$, $\xi$
being the coherence length (type-II superconductor). The length $\xi$
measures the coherence of the magnetic monopole condensate (the dual
version of the Cooper condensate). To determine the coherence length
one should measure the correlation between the chromomagnetic
monopoles. To do this one should construct a monopole creation
operator. Unfortunately, thus far there is no a convincing proposal
for the monopole operator. However, recently a promising proposal has
been advanced in Ref.~\cite{DelDebbio94}. We shall return on this
matter in Section~\ref{Section_V} For the time being, because we are
not able to determine the coherence length, we analyze our data far
from the coherence region. To this end we try a fit with the
transverse distribution~\eqref{London} by discarding the points
nearest to the flux tube ($x_t=0$).

\begin{figure}
\includegraphics[width=\textwidth]
{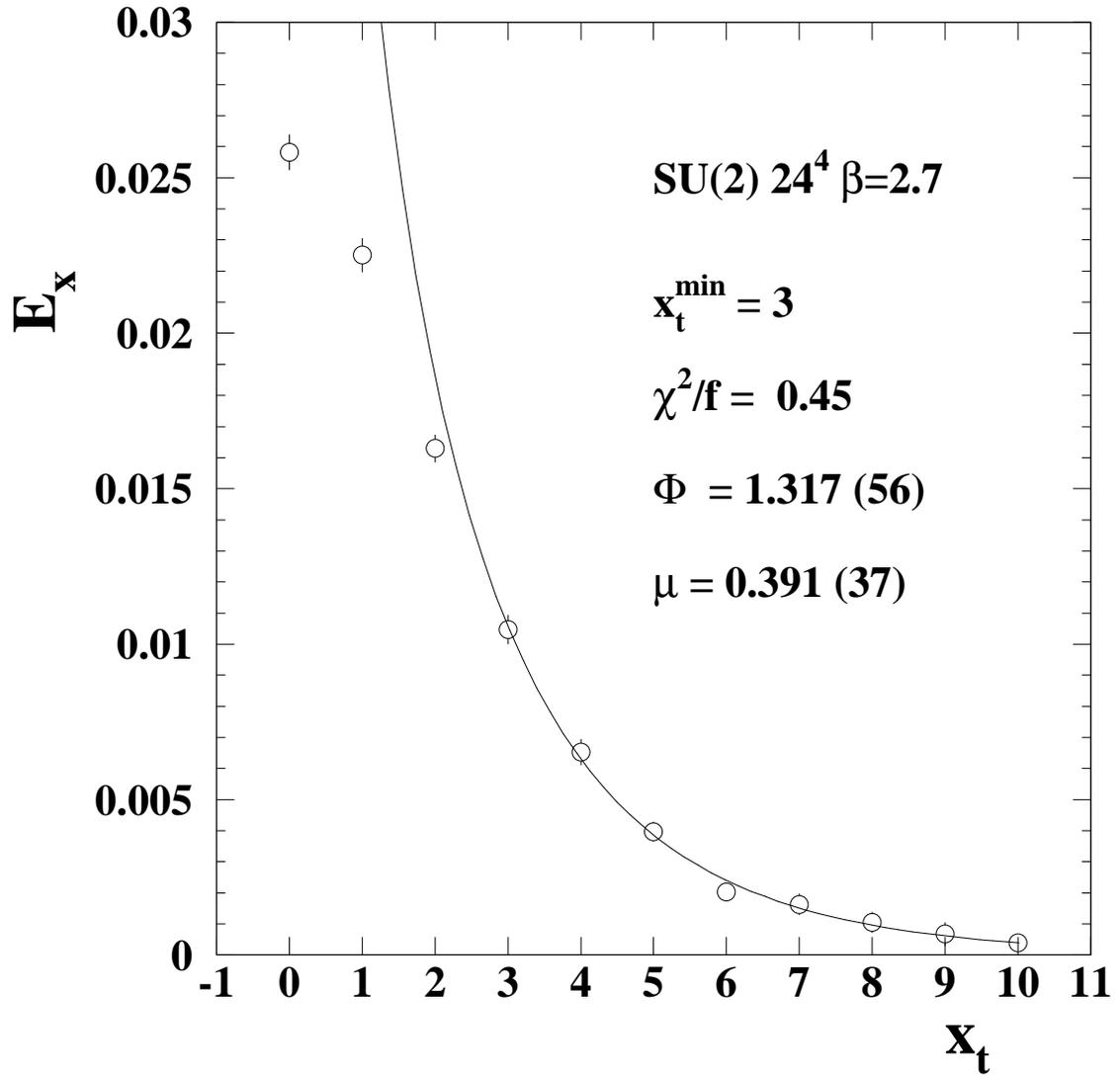}
\caption{The London fit~\protect\eqref{London} to the data for the
longitudinal chromoelectric field.}
\label{Fig:El_fit}
\end{figure}

\begin{figure}
\includegraphics[width=\textwidth]
{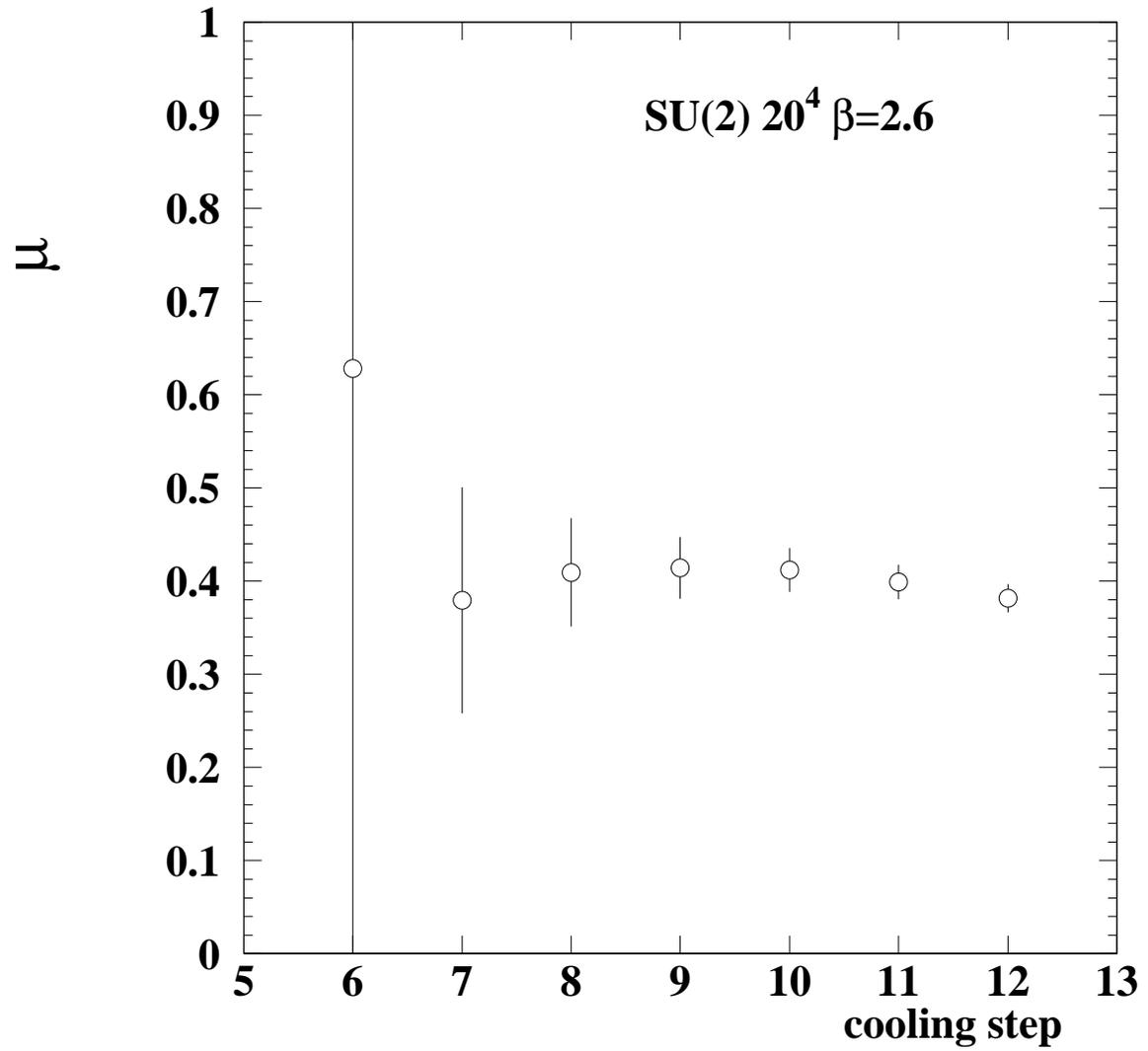} 
\caption{The inverse of the penetration length $\mu$ versus the number
of the cooling steps obtained by fitting the transverse profile of the
longitudinal chromoelectric field at $x_l=0$ ($8 \times 8$ Wilson
loop).}
\label{Fig:mu_vs_cooling}
\end{figure}

\begin{figure}
\includegraphics[width=\textwidth]
{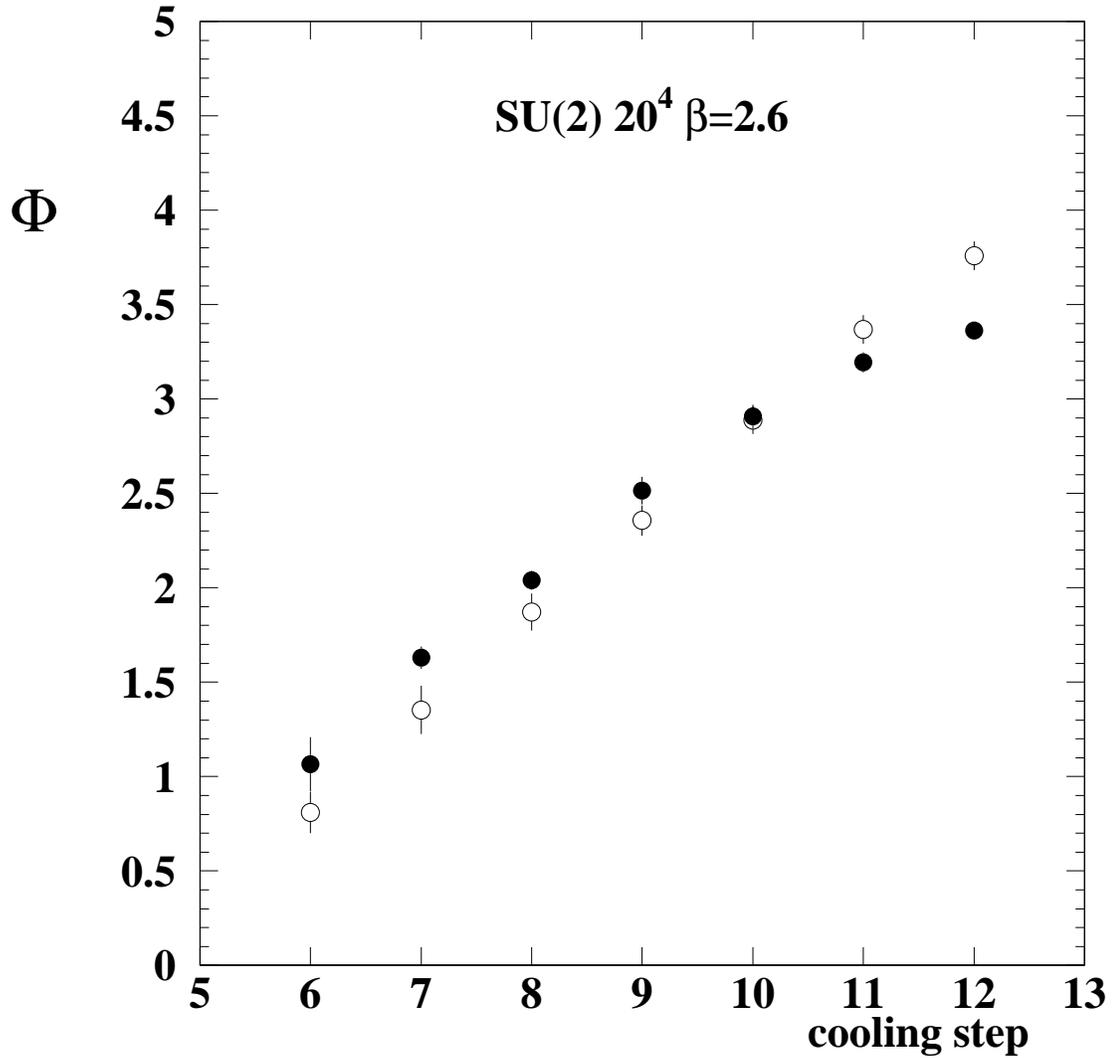}
\caption{The parameter $\Phi$ in Eq.~\protect\eqref{London} obtained
by fitting the transverse profile of the longitudinal chromoelectric
field at $x_l=0$ versus the number of cooling steps. Open points refer
to $8 \times 8$ Wilson loop, full points to $10 \times 5$ Wilson
loop.}
\label{Fig:Phi_vs_cooling}
\end{figure}

Let us discuss, firstly, the gauge invariant correlator
Eq.\eqref{rhoW}. We fit Eq.\eqref{London} to our data for $x_t \ge 2$
(in lattice units) obtaining $\chi^2/f \lesssim 1$ (we used the {\sc
Minuit} code  from the CERNLIB).
In Figure~\ref{Fig:El_fit} we show $E_l(x_t)$ measured in the middle
of the flux tube together with the result of our fit. The fit results
into the two parameters $\Phi$ and $\mu$. We have checked the
stability of these parameter by fitting Eq.\eqref{London} to the data
with the cuts $x_t \ge x_t^{\text{min}}$, $x_t^{\text{min}} =
2,3,4,5$. In Table~\ref{Table:El_fit} we report the results of our
stability analysis. We can see that within the statistical
uncertainties the fit parameters are quite stable. So we are confident
that our determination of the London penetration length is
trustworthy. We ascertained, moreover, that the data obtained from the
gauge-invariant correlator with cooled gauge configurations leads to a
parameter $\mu$ which shows a plateau versus the number of cooling
steps (see Fig.~\ref{Fig:mu_vs_cooling}).
This corroborates our expectation that the long range physics is
unaffected by the cooling procedure. On the other hand,
Figure~\ref{Fig:Phi_vs_cooling} indicates that the overall
normalization of the transverse distribution of the longitudinal
chromoelectric field is affected by the cooling. In fact the
parameter $\Phi$ does not stay constant with the cooling. We feel that
this is an indication that the flux $\Phi$ is strongly affected by
lattice artefact. This point will be thoroughly discussed below.

\begin{figure}
\includegraphics[width=\textwidth]
{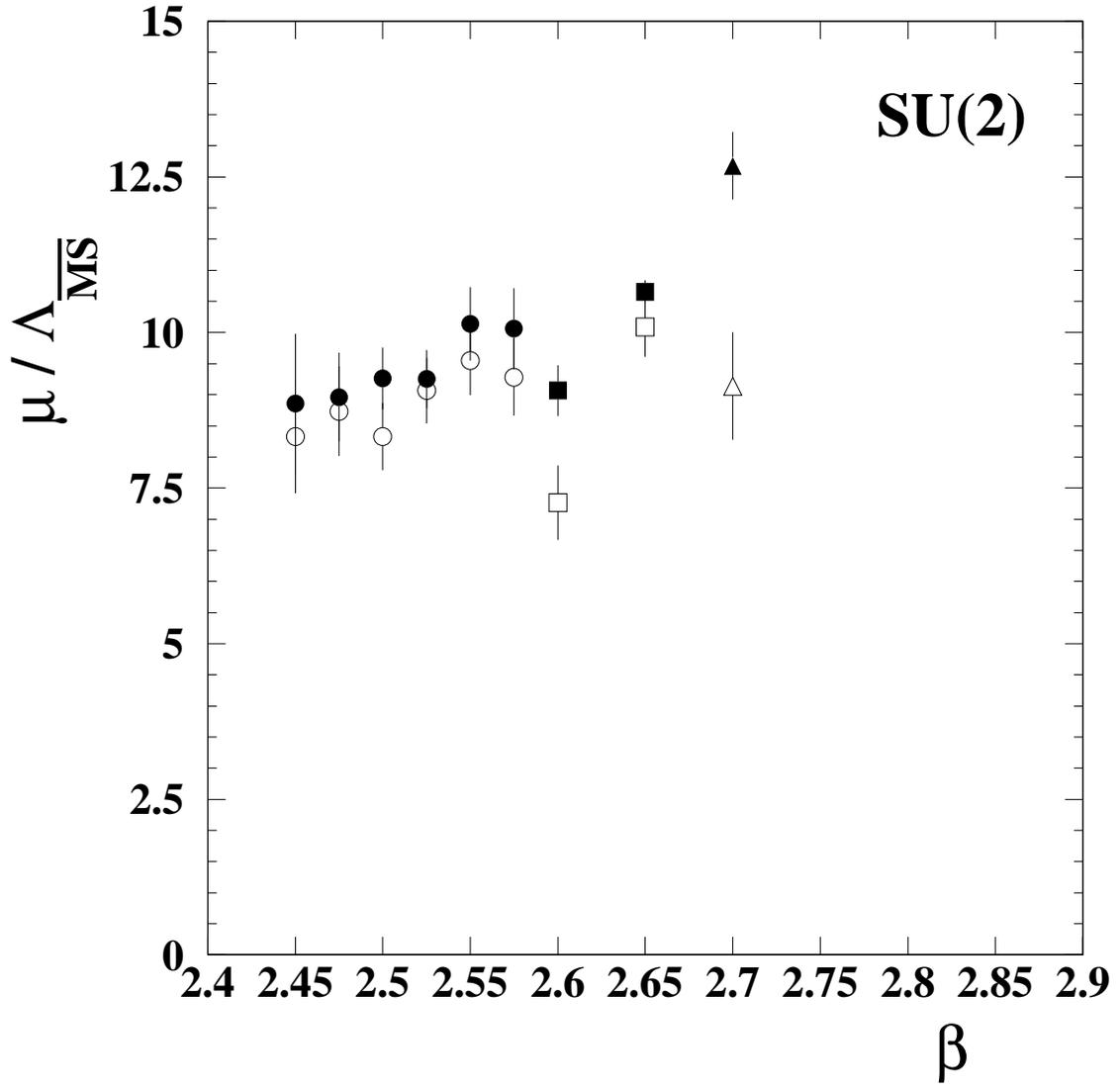}
\caption{$\mu/\Lambda_{\overline{MS}}$ versus $\beta$. Full points
correspond to rectangular Wilson loops, open points to square Wilson
loops. Circles $L=16$, squares $L=20$, and triangles $L=24$.}
\label{Fig:mu_vs_beta}
\end{figure}

\begin{figure}
\includegraphics[width=\textwidth]
{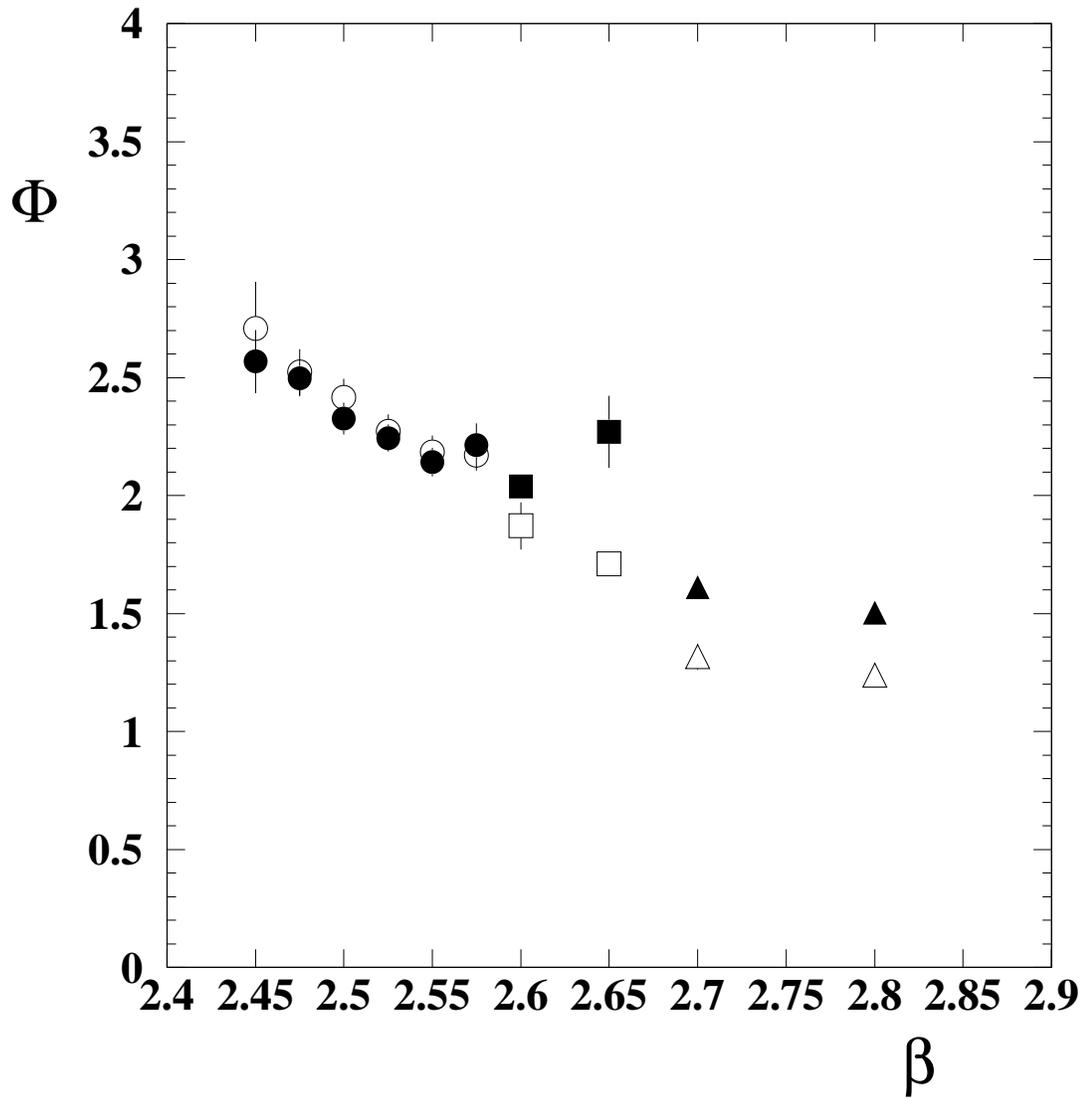}
\caption{$\Phi$ versus $\beta$. Symbols as in
Fig.~\protect{\ref{Fig:mu_vs_beta}}.} 
\label{Fig:Phi_vs_beta}
\end{figure}

In Figures~\ref{Fig:mu_vs_beta} and~\ref{Fig:Phi_vs_beta} we display
the inverse of the penetration length $\mu$ (in units of
$\Lambda_{\overline{MS}}$) and the external flux $\Phi$ versus
$\beta$. These data are obtained by fitting Eq.\eqref{London} to the
data extracted from square Wilson loop (open points) and rectangular
Wilson loops (full points). A few comments are in order. A look at
Figure~\ref{Fig:mu_vs_beta} shows that the inverse of the penetration
length $\mu$ agrees within statistical fluctuations for both kinds of
Wilson loops. However we see that for $\beta \gtrsim 2.65$ the
parameter $\mu$ arising from the rectangular Wilson loops seem to
display sizeable finite volume effects. On the other hand we find that
the parameter $\mu$ extracted from the square Wilson loops displays
finite volume effects for $\beta > 2.7$, in the case of the $24^4$
lattice. So in order to simulate in the range $\beta > 2.7$ we need
lattices with $L > 24$.

Figure~\ref{Fig:mu_vs_beta} suggests that the ratio
$\mu/\Lambda_{\overline{MS}}$ displays an approximate plateau in $\beta$.
Indeed we fitted the ratio with a constant and obtained
\begin{equation}
\label{squareWL}
\frac{\mu}{\Lambda_{\overline{MS}}} = 8.96 (31) \;, \qquad 
\chi^2/f=2.11 \,.
\end{equation}
using square Wilson loops in Eq.~\eqref{rhoW}, and
\begin{equation}
\label{rectangularWL}
\frac{\mu}{\Lambda_{\overline{MS}}} = 9.36 (29) \;, \qquad 
\chi^2/f=0.53 \,.
\end{equation}
for rectangular Wilson loops (discarding in the fit the points at
$\beta \ge 2.65$).

Equations\eqref{squareWL} and\eqref{rectangularWL} corroborates our
previous observation on the consistency of the penetration length. By
fitting all the data we obtain
\begin{equation}
\label{allWL}
\frac{\mu}{\Lambda_{\overline{MS}}} = 9.17 (21) \;, \qquad 
\chi^2/f=1.48  \,.
\end{equation}
It is worthwhile to stress that our evidence for asymptotic scaling of
the penetration length is only indicative. In general it should be
much easier to check scaling rather than asymptotic scaling. We looked
at the scaling of $\mu$ extracted from square Wilson loops with the
square root of the string tension (we have used the string tension
extracted from large Wilson loops). We found that there is approximate
scaling of $\mu$ with $\sqrt{\sigma}$ for $\beta \ge 2.5$:
\begin{equation}
\label{muoverstring}
\frac{\mu}{\sqrt{\sigma}} = 4.04 (18) \;, \qquad   \chi^2/f=1.38 \,.
\end{equation}
So we see that our data on the penetration length are in agreement
with the general expectation that scaling goes better than asymptotic
scaling. On the other hand, the approximate evidence of asymptotic
scaling is a natural consequence of the fact that the penetration
length is a physical quantity related to the size $D$ of the flux
tube~\cite{Cea93}:
\begin{equation}
\label{fluxtubesize}
D  \simeq  \frac{2}{\mu} \;.
\end{equation}
As concern the parameter $\Phi$, Figure~\ref{Fig:Phi_vs_beta} shows
that $\Phi$ is rather insensitive to the shape of the Wilson loops
used in Eq.\eqref{rhoW} (again the data from rectangular Wilson loops
are affected by finite volume effects for $\beta \gtrsim 2.65$).
Moreover $\Phi$  decreases rapidly by increasing 
$\beta$ and seems to saturate to a value quite close to 1. We postpone
the discussion of this behaviour until the comparison with the results
obtained using Abelian projected configurations in the maximally
Abelian gauge.

\renewcommand{\thesubsection}{\Alph{subsection}.}
\subsection{Maximally Abelian projection}
\label{Section_III_B}

\begin{figure}
\includegraphics[width=\textwidth]
{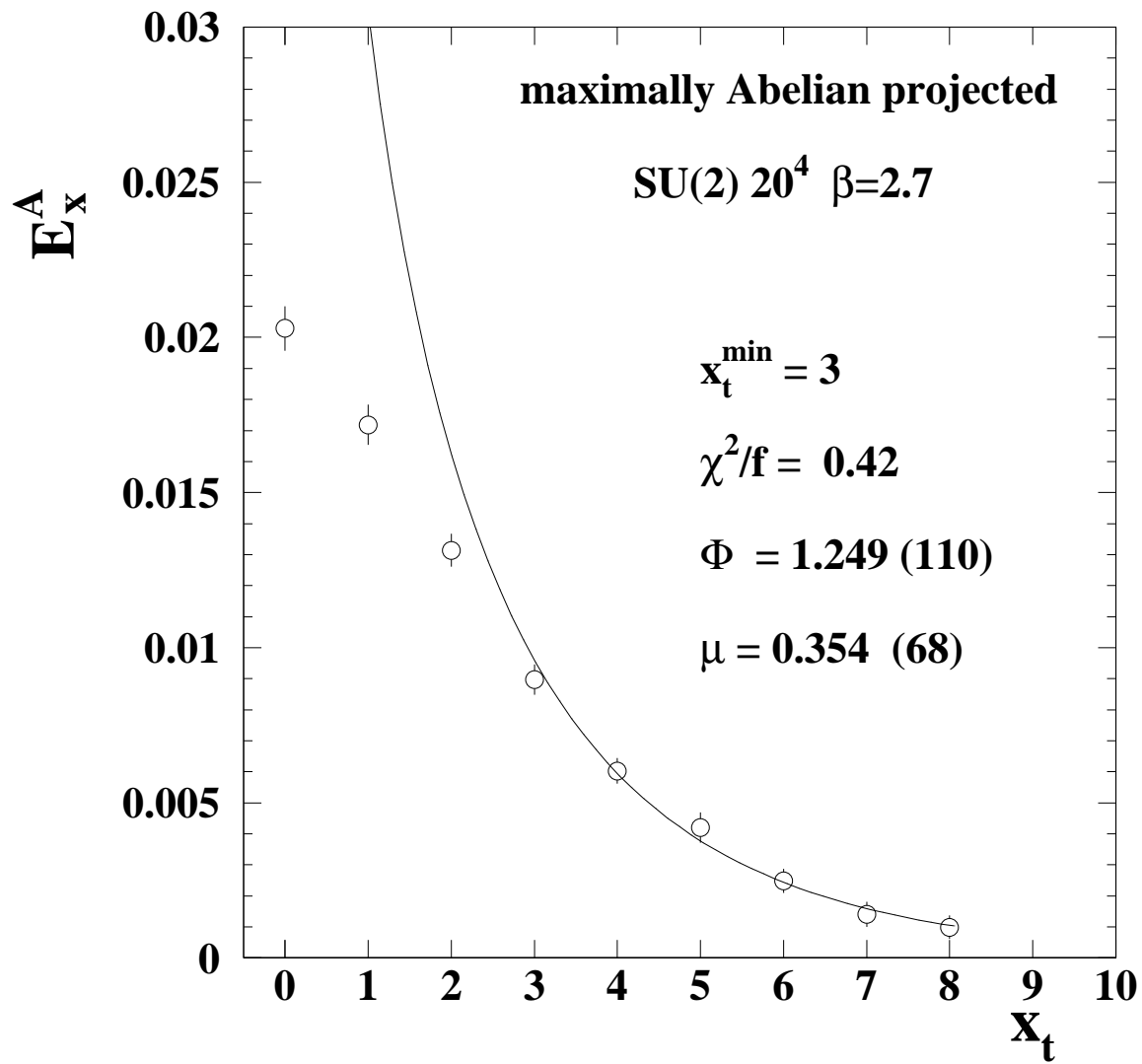}
\caption{London fit~\protect\eqref{LondonAb} to the data for the
Abelian  longitudinal chromoelectric field at $x_l=0$ for square
Wilson loop.}
\label{Fig:El_Ab_fit}
\end{figure}

\begin{figure}
\includegraphics[width=\textwidth]
{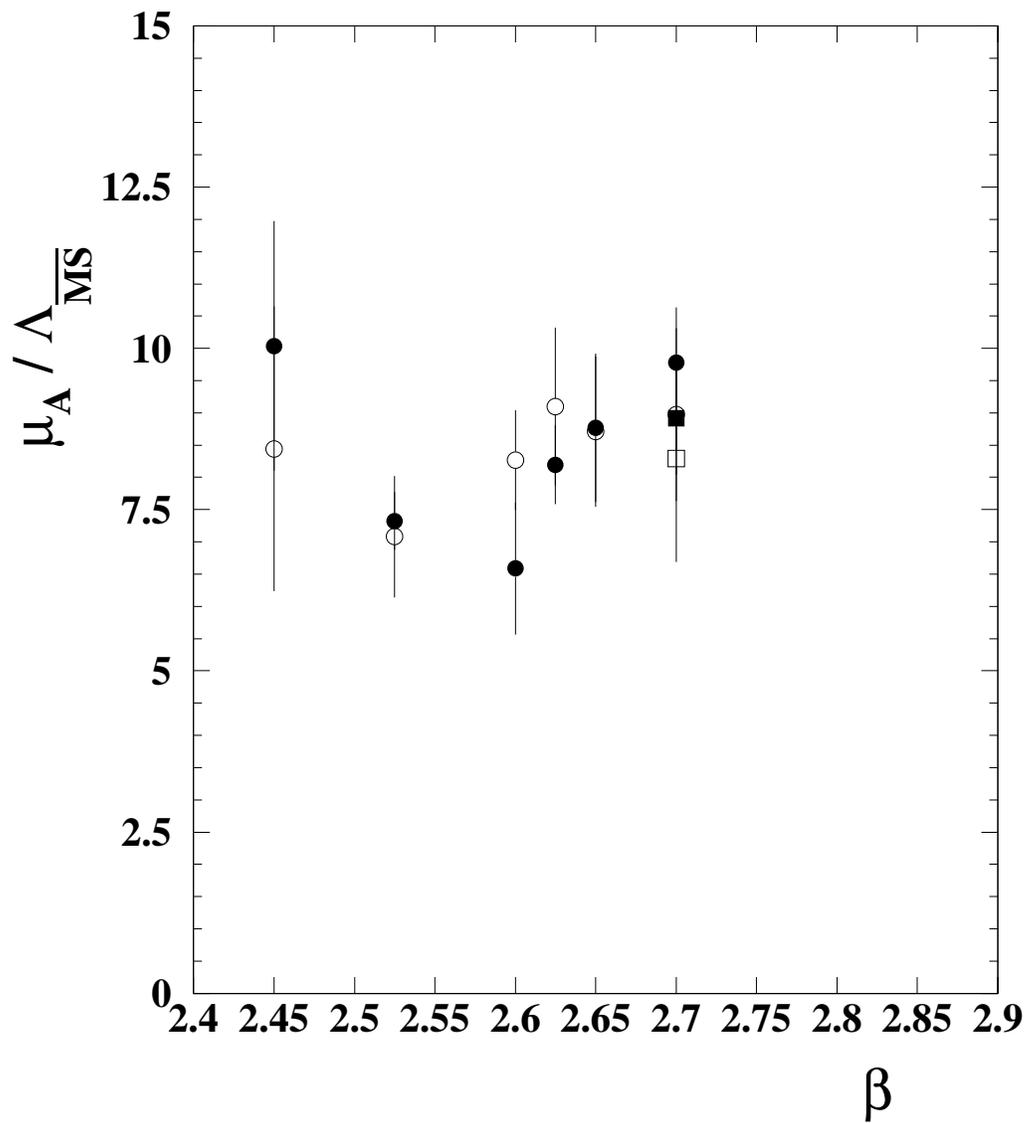}
\caption{$\mu_A/\Lambda_{\overline{MS}}$ versus $\beta$. Full points
correspond to rectangular Wilson loops, open points to square Wilson
loops. Circles $L=16$, squares $L=20$.}
\label{Fig:muAb_vs_beta}
\end{figure}

Let us consider, now, the Abelian projected field strength
tensor Eq.~\eqref{FmunuAb}. As we saw, only the longitudinal Abelian
chromoelectric field is sizeable. As in previous case we try to fit
the data with the law:
\begin{equation}
\label{LondonAb}
E^A_l(x_t) = \frac{\Phi_A}{2 \pi} \mu_A^2 K_0(\mu_A x_t) \;, x_t > 0
\;.
\end{equation}
Again we find (see Fig.~\ref{Fig:El_Ab_fit}) that Eq.~\eqref{LondonAb}
reproduces quite well our data for $x_t \ge 2$ ($\chi^2/f \lesssim
1$). In Table~\ref{Table:ElAb_fit} we check the stability of the fit
parameters. In Figure~\ref{Fig:muAb_vs_beta} we display the ratio
$\mu_A/\Lambda_{\overline{MS}}$ obtained by fitting
Eq.\eqref{LondonAb} to the data in the case of square Wilson loops
(open points) and rectangular Wilson loops (full points). Within the
(rather large) statistical uncertainties the parameter $\mu_A$ agrees
for the two different Wilson loops. Moreover the data suggest that the
ratio $\mu_A/\Lambda_{\overline{MS}}$ does not depend on $\beta$.
Indeed we fit the ratio with a constant and find 
\begin{equation}
\label{squareWLAb}
\frac{\mu_A}{\Lambda_{\overline{MS}}} = 8.26 (67) \;, \qquad 
\chi^2/f=0.41 \,. 
\end{equation}
using square Wilson loops in Eq.~\eqref{rhoWab}, and
\begin{equation}
\label{rectangularWLAb}
\frac{\mu_A}{\Lambda_{\overline{MS}}} = 8.27 (52) \;, \qquad 
\chi^2/f=1.87 \,.
\end{equation}
for rectangular Wilson loops. An overall fit of all the data gives
\begin{equation}
\label{allWLAb}
\frac{\mu_A}{\Lambda_{\overline{MS}}} = 8.27 (41) \;, \qquad 
\chi^2/f=1.05 \;. 
\end{equation}
Note that Eq.~\eqref{squareWL} and
Eqs.~\eqref{squareWLAb}-\eqref{allWLAb} give consistent value for the
ratio $\mu/\Lambda_{\overline{MS}}$. On the other hand the ratio
$\mu/\Lambda_{\overline{MS}}$, Eq.~\eqref{rectangularWL}, extracted
from the gauge invariant correlator $\rho_W$ with rectangular Wilson
loops is slightly higher than Eqs.~\eqref{squareWLAb}-\eqref{allWLAb}.
Indeed, Eq.~\eqref{rectangularWL} and
Eqs.~\eqref{squareWLAb}-\eqref{allWLAb} are consistent within two
standard deviations. We feel that this small discrepancy is due to the
fact that the rectangular Wilson loops seem to be more sensitive to
finite volume effects. For this reason we shall, henceforth, refer to
the data extracted from square Wilson loops.

\begin{figure}
\includegraphics[width=\textwidth]
{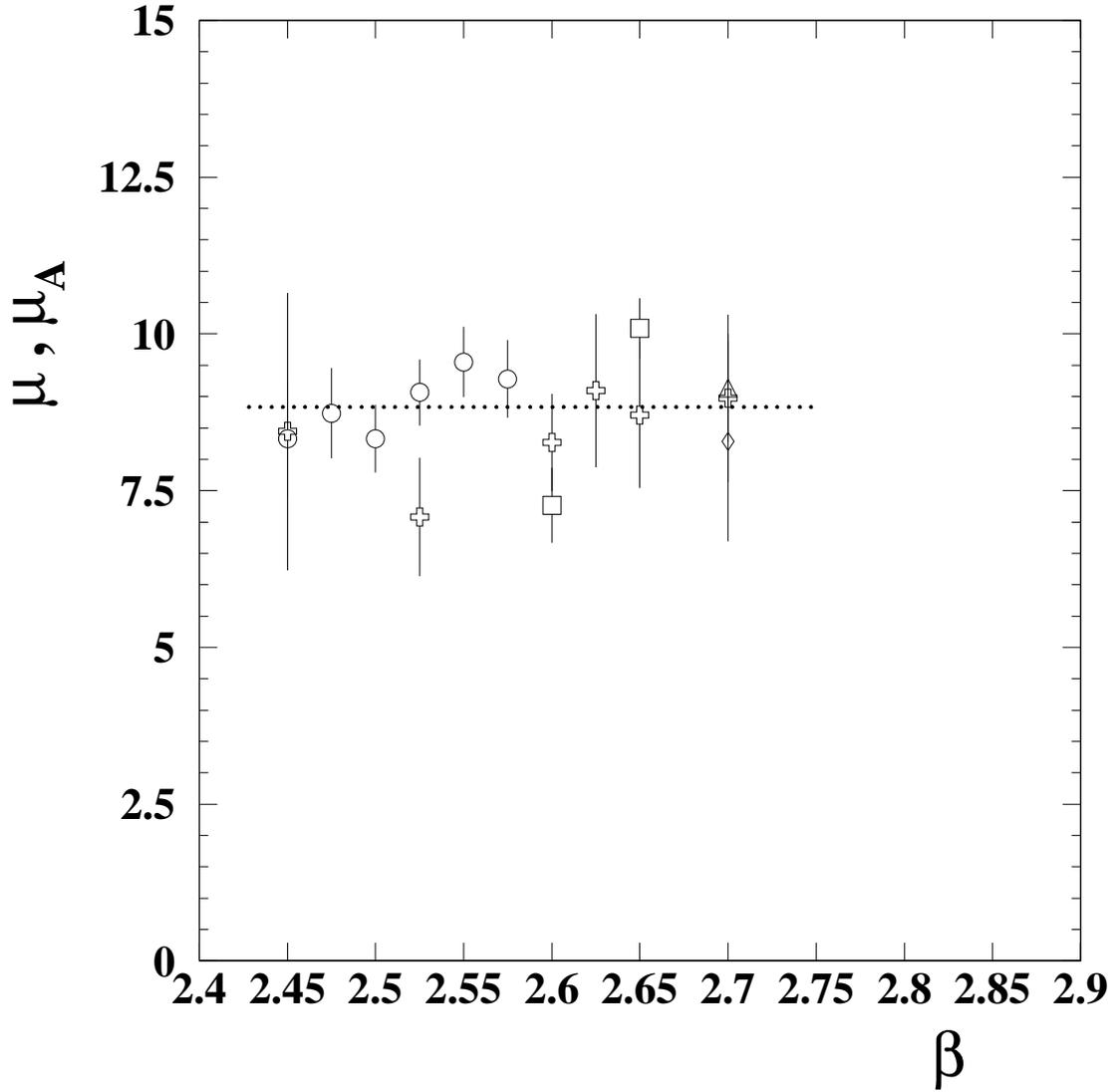}
\caption{$\mu$ and $\mu_A$ (in units of $\Lambda_{\overline{MS}}$)
versus $\beta$  for square Wilson loops. Circles, squares, and
triangle refer to $L=16$, $20$, $24$ respectively. Crosses and diamond
refer to the Abelian projected correlator $\rho^A_W$ with $L=16$, $20$
respectively.}
\label{Fig:mu_vs_beta_Ab-nonAb}
\end{figure}

In Fig.~\ref{Fig:mu_vs_beta_Ab-nonAb} we report the ratio
$\mu/\Lambda_{\overline{MS}}$ and $\mu_A/\Lambda_{\overline{MS}}$
versus $\beta$ obtained by the data corresponding to
square Wilson loops. We can see that the London
penetration length extracted from the gauge-invariant correlator
Eq.~\eqref{rhoW} agrees with the one extracted from the Abelian
projected correlator Eq.~\eqref{rhoWab}. 
In Figure~\ref{Fig:mu_vs_beta_Ab-nonAb} we show also the result
obtained by fitting together the data (for square Wilson loops):
\begin{equation}
\label{mualldata}
\frac{\mu}{\Lambda_{\overline{MS}}} = 8.84 (28) \;, \qquad 
\chi^2/f=1.44 \;. 
\end{equation}
As a consequence we can safely affirm that the London penetration
length is gauge invariant. We feel that this results strongly supports
the dual superconductor mechanism of confinement. 

\begin{figure}
\includegraphics[width=\textwidth]
{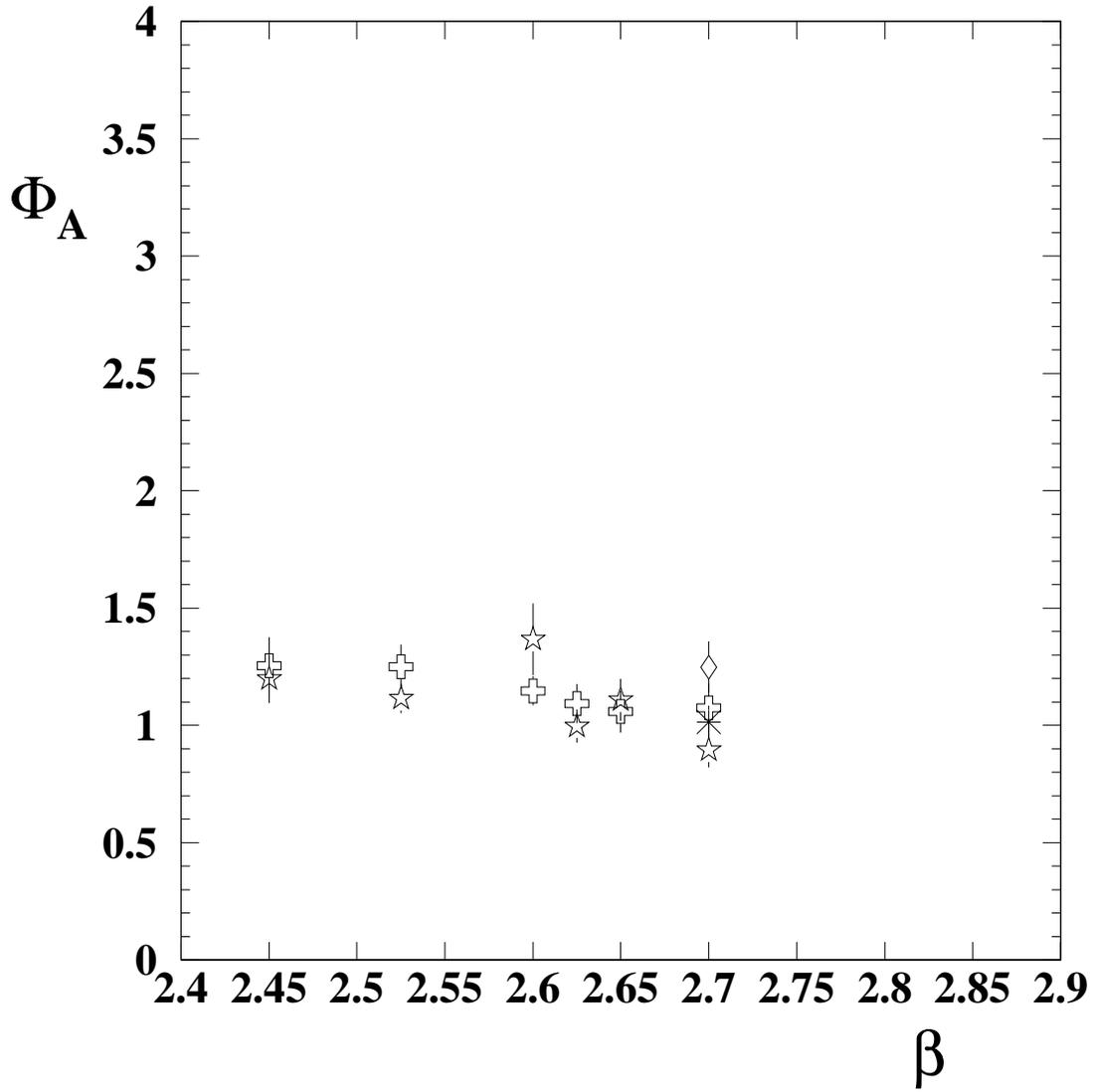}
\caption{$\Phi_A$ versus $\beta$. Crosses and diamond correspond to
square Wilson loops with $L=16$, $20$ respectively; stars and
asterisk correspond to rectangular Wilson loop with $L=16$ and $20$
respectively.} 
\label{Fig:Phi_vs_beta_Ab}
\end{figure}

As concern the parameter $\Phi_A$, we find that, unlike the previous
case, $\Phi_A$ does not depend strongly on $\beta$ (see
Fig.~\ref{Fig:Phi_vs_beta_Ab}). Moreover we see that $\Phi_A$ is
quite close to 1. It is worthwhile to discuss the physical
interpretation of $\Phi$. The total flux $\Phi_T$ of the flux tube
chromoelectric field is given by
\begin{equation}
\label{totalflux}
\Phi_T = \int d^2 x_t E_l(x_t) \;,
\end{equation}
where the integral extends over a plane transverse to the line joining
the static color charges. As we have already discussed, the transverse
distribution of the longitudinal chromoelectric field can be described
by the law Eq.~\eqref{London} when $x_t > 0$. Obviously we cannot
extend the validity of Eq.~\eqref{London} up to $x_t \rightarrow 0$.
Indeed for $x_t \rightarrow 0$ we encounter a logarithmic divergence
in $K_0$. On the other hand, $E_l(x_t)$ is finite in the coherence
region $x_t \lesssim \xi$. However, if $\lambda/\xi \gtrsim 1$ we
extimate that the extrapolation up to the origin introduces an
overextimation of the integral~\eqref{totalflux} by less than $10\%$.
So, inserting~\eqref{London} into~\eqref{totalflux}, we get:
\begin{equation}
\label{approxtotalflux}
\Phi_T = \int d^2 x_t E_l(x_t) \simeq \Phi   \;.
\end{equation}
Equations~\eqref{totalflux} and~\eqref{approxtotalflux} tell us that
the parameter $\Phi$ measures the total flux if $\lambda/\xi \gg 1$.
In U(1) it turns out that $\Phi=1$, since that happens to be one unit
of quantized electric flux~\cite{Singh93}. If the dynamics of the
Abelian projected fields resembles the gauge fields of U(1), then we
expect that $\Phi_A \simeq 1$. Indeed we find (square Wilson loops)
\begin{equation}
\label{approxPhiAb}
\Phi_A = 1.15 (5) \,, \qquad \chi^2/f = 0.79 \,.
\end{equation}
From the previous discussion it follows that Eq.~\eqref{approxPhiAb}
seems to indicate that $\lambda/\xi \sim 1$.

\begin{figure}
\includegraphics[width=\textwidth]
{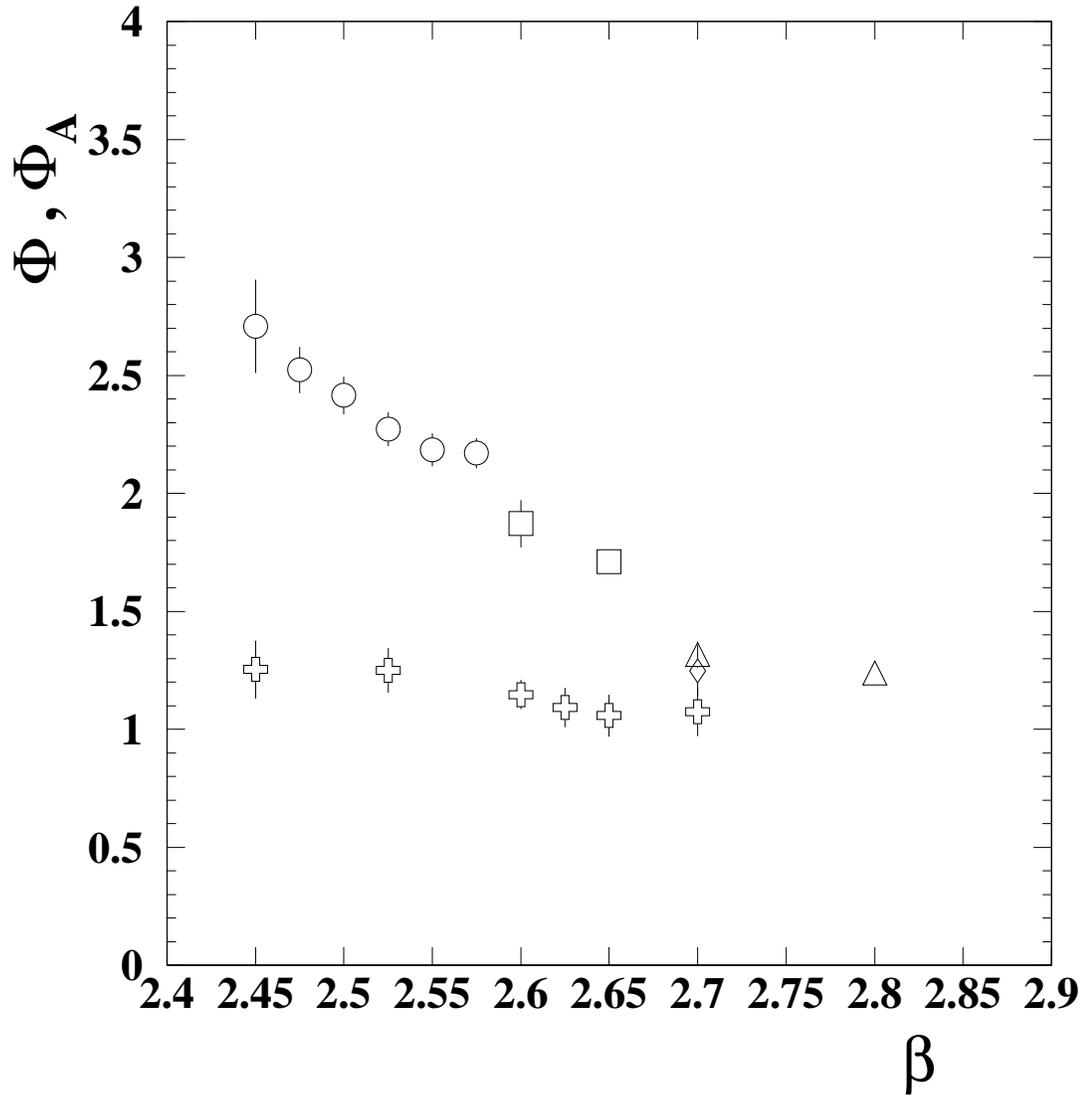}
\caption{$\Phi$ and $\Phi_A$ versus $\beta$ for square Wilson loops.
Points and crosses refer to $L=16$, squares and diamond to $L=20$,
triangles to $L=24$. Crosses and diamond correspond to the maximally
Abelian gauge.}
\label{Fig:Phi_vs_beta_Ab-nonAb}                                    
\end{figure}

We would like to contrast Eq.~\eqref{approxPhiAb} with the behaviour
of $\Phi$. In Figure~\ref{Fig:Phi_vs_beta_Ab-nonAb} we report $\Phi_A$
and $\Phi$ versus $\beta$. The behavior of $\Phi$ under the cooling
(see Fig.~\ref{Fig:Phi_vs_cooling}) suggested that the external flux
is strongly affected by lattice artefacts. Moreover,
Figure~\ref{Fig:Phi_vs_beta_Ab-nonAb} indicates that the lattice
artefacts  seem to disappear by increasing $\beta$. Thus we are led to
suspect that the external flux gets renormalized by irrelevant
operators, whose effects are strongly suppressed in the maximally
Abelian gauge.

\renewcommand{\thesection}{\Roman{section}.}
\setcounter{equation}{0}
\section{STRING TENSION}
\label{Section_IV}
\renewcommand{\thesection}{\arabic{section}}

In the previous Section we have shown that the color fields of a
static quark-antiquark pair are almost completely described by the
longitudinal chromoelectric field. Moreover we showed that the
longitudinal chromoelectric field is almost constant along the flux
tube. This means that the long distance potential which feels the
color charges is linear. Obviously the string tension is given by the
energy stored into the flux tube per unit length. As a consequence we
can write
\begin{equation}
\label{string_tension}
\sigma \simeq \frac{1}{2} \int d^2 x_t E_l^2(x_l,x_t) \;.
\end{equation}
We stress that the string tension $\sigma$ defined by
Eq.~\eqref{string_tension} does not depend on $x_l$ as long as the
longitudinal chromoelectric field is constant along the flux tube. As
we have already discussed, working on a finite lattice results in the
limitations $x_l=0,\pm 1$ (in lattice units) in the integrand in
Eq.~\eqref{string_tension}. Keeping these limitations in mind, from
Equation~\eqref{string_tension} we can obtain an explicit relation
between the string tension and the parameters $\Phi$ and $\mu$.
Indeed, if we extrapolate Eq.~\eqref{London} up to $x_t=0$, by using 
\begin{equation}
\label{K0integral}
\int_0^\infty dx \, x K_0^2(x) = \frac{1}{2}  \;,
\end{equation}
we get
\begin{equation}
\label{sqrtstring}
\sqrt{\sigma} \simeq \frac{\Phi}{\sqrt{8 \pi}}  \mu  \;.
\end{equation}
The main uncertainty in Eq.~\eqref{sqrtstring} comes out from the
parameter $\Phi$. As explained in Sects.~II  and~III we computed the
parameters $\Phi$ and $\mu$ on SU(2) gauge configurations and on the
maximally Abelian projected gauge configurations. In the latter case
$\Phi_A \thickapprox 1$ and independent of $\beta$. On the other hand,
for SU(2), $\Phi > 1$ and it approaches values very close to $\Phi_A$
by increasing $\beta$. As we have already discussed, we feel that the
external flux $\Phi$ is strongly affected by lattice artefacts. We can
try to get rid of these effects by assuming that in the limit $\beta
\rightarrow \infty$ 
\begin{equation}
\label{PhieqPhiAb}
\Phi \simeq  \Phi_A \simeq  1 \;.
\end{equation}
In this way Eq.~\eqref{sqrtstring} becomes
\begin{equation}
\label{sqrtstringdef}
\sqrt{\sigma} \simeq \frac{\mu}{\sqrt{8 \pi}} \;.
\end{equation}
A striking consequence of Eq.~\eqref{sqrtstringdef} is that, due to
$\mu \simeq \mu_A$,
\begin{equation}
\label{abstreqstr}
\sqrt{\sigma} \simeq \sqrt{\sigma}_A
\end{equation}
within statistical uncertainties.

\begin{figure}
\includegraphics[width=\textwidth]
{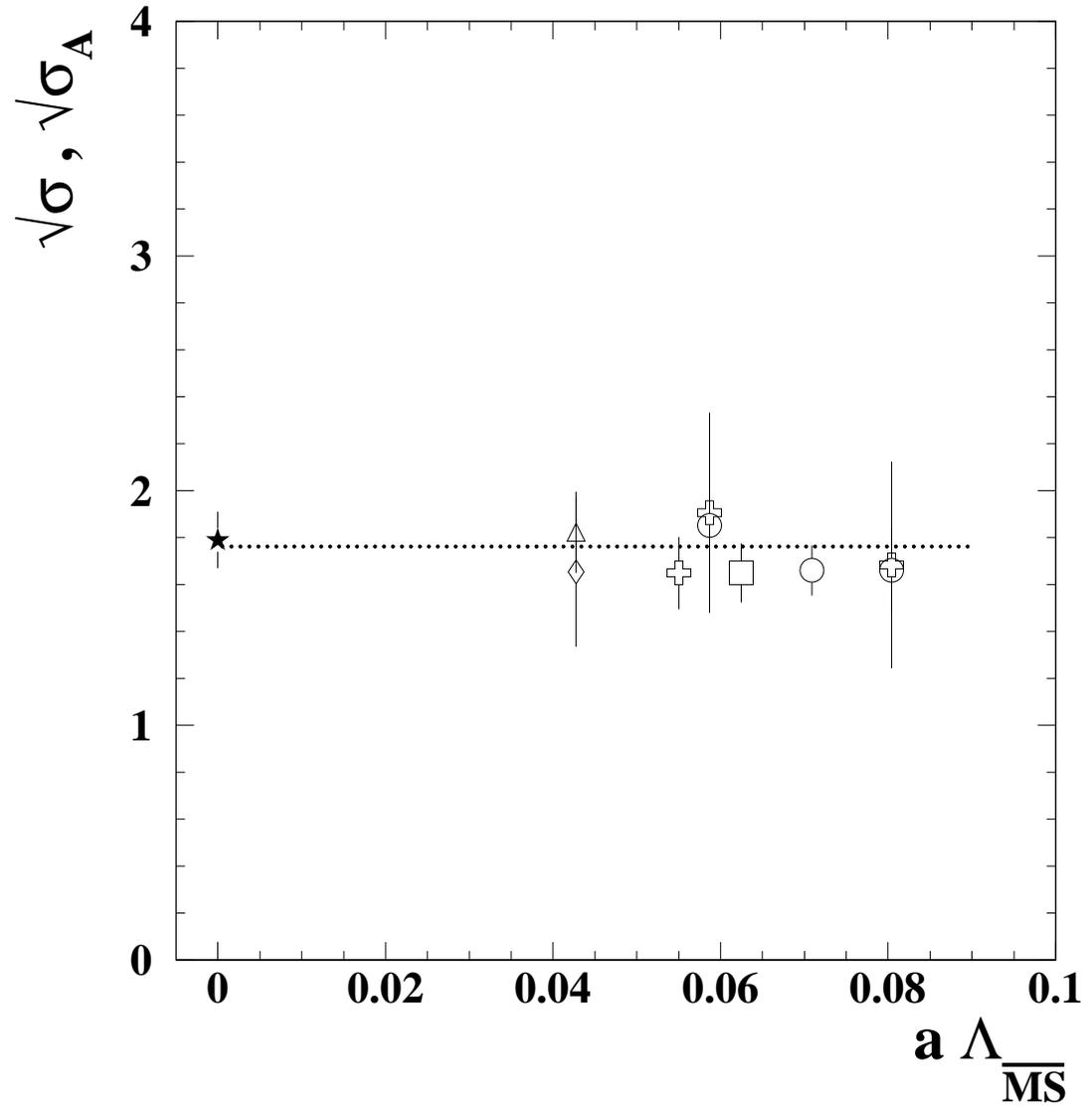}
\caption{String tension (in units of $\Lambda_{\overline{MS}}$)
evaluated through
Eq.~\protect\eqref{string_tension}. Star refers to the value given in
Ref.~\protect\cite{Fingberg93}. Symbols as in
Fig.~\protect{\ref{Fig:Phi_vs_beta_Ab-nonAb}}. For figure readability
not all the available data are displayed.}
\label{Fig:string_tension}
\end{figure}

In Figure~\ref{Fig:string_tension} we report Eq.~\eqref{sqrtstringdef}
in units of $\Lambda_{\overline{MS}}$ versus $a
\Lambda_{\overline{MS}}$. Fitting all together the data to a constant
we get (square Wilson loops)
\begin{equation}
\label{stringvalue}
\frac{\sqrt{\sigma}}{\Lambda_{\overline{MS}}} = 1.76(6) \,, \qquad
\chi^2/f=1.44  \,.
\end{equation}
The quoted error in Eq.~\eqref{stringvalue} is purely statistic.
However, one should keep in mind that our theoretical uncertainties in
the extimation of the string tension~\eqref{stringvalue} introduce a
systematic error which can be of the order of ten per cent.
Nevertheless, it is gratifying to see that our extimation of the
string tension Eq.~\eqref{stringvalue} is  consistent with (star in
Fig.~\ref{Fig:string_tension})
\begin{equation}
\label{wilsonstring}
\frac{\sqrt{\sigma}}{\Lambda_{\overline{MS}}} = 1.79(12) \;.
\end{equation}
The value quoted in Eq.~\eqref{wilsonstring} has been obtained in
Ref.~\cite{Fingberg93} by the linear asymptotic extrapolation of the
string tension data extracted from Wilson loops on lattices larger
than ours.

\renewcommand{\thesection}{\Roman{section}.}
\setcounter{equation}{0}
\section{CONCLUSIONS}
\label{Section_V}
\renewcommand{\thesection}{\arabic{section}}

Let us conclude by stressing the main results of this paper. We
investigated the color field strength tensor of the $q-\bar{q}$ flux
tube by means of the connected correlators~\eqref{rhoW} (full SU(2))
and~\eqref{rhoWab} (maximally Abelian gauge). 

The main advantage of using the connected correlator~\eqref{rhoW} 
and~\eqref{rhoWab} resides in the fact that the connected correlators
are sensitive to the field strength rather than to the square of the
field strength. As a consequence we are able to detect a sizeable
signal even with relatively low statistics. It turns out that the flux
tube color fields is composed by the chromoelectric component parallel
to the line joining the static charges. Moreover the longitudinal
chromoelectric field is almost constant far from the color sources,
and it decreases rapidly in the directions transverse to the line
connecting the charges. As a matter of fact we found that the
transverse distribution of the longitudinal chromoelectric field
behaves in accord with the dual Meissner effect. This allows us to
determine the London penetration length. We checked that the
penetration length is a physical gauge invariant quantity. A
remarkable consequence of our findings is that the long range
properties of the SU(2) confining vacuum can be described by an
effective Abelian theory. In addition, after fixing the gauge with the
constraints~\eqref{DmuAmu}-\eqref{R}, it seems that the degrees of
freedom which are not relevant to the confinement get suppressed. 

Finally, we put out a very simple relation between the string tension
and the penetration length which gives an extimate of
$\sqrt{\sigma}$ quite close to the extrapolated continuum limit
available in the literature.

In conclusion we would like to stress that the most urgent problem to
be addressed in the future studies is the reliable extimation of the
coherence length $\xi$. The results in Sect.~III give an indirect and,
admittedly,  very weak indication that $\lambda/\xi \sim 1$.  As we
have already discussed, the coherence length is determined by the
monopole condensate, the order parameter for the confinement. Recently
two different groups~\cite{Haymaker93,Matsubara94} give an extimation
of the coherence length. These authors calculate the electric flux and
magnetic monopole current distribution in the presence of a static
quark-antiquark pair for SU(2) lattice gauge theory in the maximally
Abelian gauge. The magnetic monopoles are identified using the
DeGrand-Toussaint~\cite{DeGrand80} construction. By using a dual form
of the Ginzburg-Landau theory~\cite{Ginzburg-Landau}, which allows the
magnitude of the monopole condensate density to vary in space, they
fitted the data and obtain $\lambda/a$ and $\xi/a$. They found that
the coherence length is comparable to the penetration length. Even
though we feel that the approach of
Refs.~\cite{Haymaker93,Matsubara94} is interesting, we would like to
observe that it relies heavily on the definition of the magnetic
monopole current. As a matter of fact, in Ref.~\cite{DelDebbio91} it
was pointed out that the DeGrand-Toussaint definition of the monopole
density is plagued by lattice artefacts, which are, however, less
severe in maximally Abelian gauge. So that the DeGrand-Toussaint
monopole density is not an order parameter for confinement. Thus the
approach of Refs.~\cite{Haymaker93,Matsubara94} is plagued by the
ambiguities related to the definition of the monopole current. On the
other hand, in our approach we work outside the coherence region, so
that we feel that our results do not manifest the above mentioned
problem. To clarify this point, it should be of great help the study
of the distribution of color fields in the presence of a static
quark-antiquark pair in the framework of the dual Ginzburg-Landau
model with the magnetic monopole current constructed by means of the
monopole creation operator proposed in Ref.~\cite{DelDebbio94}.

\setcounter{equation}{0}
\renewcommand{\theequation}{A\arabic{equation}}
\section*{APPENDIX}
\label{Appendix}

In this Appendix we give more details on the algorithm used to fix
the maximally Abelian gauge. On the lattice, the maximally Abelian
gauge is obtained by maximizing the lattice functional 
\begin{equation}
\label{Rl}
R_l = \sum_{x,\hat{\mu}} \frac{1}{2} \text{tr} \left[ \sigma_3 U_\mu(x)
\sigma_3 U^\dagger_\mu(x) \right]
\end{equation}
over all SU(2) gauge transformations 
\begin{equation}
\label{gaugetrf}
U_\mu(x) \rightarrow \widetilde{U}_\mu(x) = g(x) U_\mu(x) g^\dagger(x)
\;,
\end{equation}
where $g(x) \, \in \,$~SU(2).
Under an arbitrary gauge transformation the variation of the
lattice functional Eq.~\eqref{Rl} is
\begin{equation}
\label{DeltaRl}
\Delta R_l(x) = \frac{1}{2} \text{tr} \left[ g^\dagger(x) \sigma_3
g(x) X(x) \right] - \frac{1}{2} \text{tr} \left[ \sigma_3 X(x) \right]
\;,
\end{equation}
where
\begin{equation}
\label{X}
X(x) = \sum_\mu \left[ U_\mu(x) \sigma_3 U^\dagger_\mu(x) +
U^\dagger_\mu(x-\hat{\mu}) \sigma_3 U_\mu(x-\hat{\mu}) \right] 
\end{equation}
belongs to the SU(2) algebra.
If we have locally maximized the lattice functional~\eqref{Rl} with
respect to an arbitrary gauge transformation, then we have
\begin{equation}
\label{DeltaRl=0}
\Delta R_l(x) = 0 \;.
\end{equation}
From Eq.~\eqref{DeltaRl} it follows that 
\begin{equation}
\label{Xdiag}
X(x) = g(x) X(x) g^\dagger(x) \;,
\end{equation}
i.e. $X(x)$ must be diagonal. So that maximizing $R_l(x)$ is
equivalent to diagonalizing the hermitian matrix $X(x)$. Note that
maximization of $R_l(x)$ at the given lattice site $x$ is accomplished
by a gauge transformation $g(x)$, which, in turn, affects the value of
the local operator $X(x)$ at the nearest neighbours. Therefore the
maximization of the lattice functional~\eqref{Rl} can be achieved only
by an iterative procedure over the whole lattice. In equivalent manner
one can find $g(x)$ as the matrix which diagonalizes $X(x)$ or as the
matrix which maximizes $R_l(x)$.

To obtain explicitly the gauge element $g(x)$ which maximizes
$R_l(x)$, let us write $R_l(x)$ as
\begin{equation}
\label{R_l(x)}
\begin{split}
R_l(x) &=   \frac{1}{2} \text{tr} \left[\sigma_3  g^\dagger(x) \sigma_3
g(x) X(x) \sigma_3 \right]\\
&= k(x) \frac{1}{2}  \text{tr} \left[\sigma_3  g^\dagger(x) \sigma_3
g(x) V(x) \right]
\end{split}
\end{equation}
where
\begin{equation}
\label{V(x)}
V(x) = \frac{X(x) \sigma_3}{k(x)} \;, \qquad  
k(x) = \sqrt{\text{det}\left(X(x) \sigma_3 \right)}
\end{equation}
ensuring that $V(x)$ is an element of SU(2).  As one can easily
recognize from Eq.~\eqref{X}
\begin{equation}
\label{V(x)-par}
V(x)= v_0(x) + i(v_1(x) \sigma_1 + v_2(x) \sigma_2)  \,.
\end{equation}
Note that the term proportional to $\sigma_3$ is absent. Now, we
observe that, if we consider $\widetilde{g}(x)=u(x) g(x)$ (with
$u(x)=u_0(x)+i u_3(x) \sigma_3$) instead of $g(x)$, then
Eq.~\eqref{R_l(x)} is invariant. So we can assume without loss of
generality that in Eq.~\eqref{R_l(x)} $g(x)=g_0(x)+i(g_1(x) \sigma_1 +
g_2(x) \sigma_2)$. As a consequence, Eq.~\eqref{R_l(x)} is maximized
when
\begin{equation}
\label{g0-g1-g2}
\begin{split}
g_0(x) &= \pm \sqrt{\frac{v_0(x)+1}{2}} \,,\\
g_1(x) &= - \frac{v_1(x)}{2 g_0(x)} \,,\\
g_2(x) &= - \frac{v_2(x)}{2 g_0(x)} \,. \\
\end{split}
\end{equation}
Since maximization of~\eqref{Rl} results in an iterative procedure, we
must have at disposal a convergence criterion. To have a measure of
the goodness of gauge fixing we consider the average size of the
non-diagonal matrix elements of $X$ over the whole lattice:
\begin{equation}
\label{off-diag}
\left\langle \left| X^{\text{nd}} \right|^2 \right\rangle =
\frac{1}{L^4} \sum_x \left[   \left| X_1 \right|^2 +  
\left| X_2 \right|^2   \right]
\end{equation}
where $X=X_1 \sigma_1 + X_1 \sigma_2 +  X_1 \sigma_3$.
We stop the iterations when
\begin{equation}
\label{criterion}
\left\langle \left| X^{\text{nd}} \right|^2 \right\rangle \le D
\end{equation}
where $D$ is some (small) positive number. In our simulation we used
$D=10^{-6}$.

In order to accelerate the convergence of the algorithm we adopted the
overrelaxation method~\cite{Adler88}  suggested in
Ref.~\cite{Mandula90}. Once we have found the matrix $g(x)$ which
maximizes~\eqref{DeltaRl}, we make the following substitution
\begin{equation}
\label{overrelaxation}
g(x) \rightarrow g_{\text{over}}(x) = g(x)^\omega
\end{equation}
where the overrelaxation parameter $\omega$ varies in the interval $1
\le \omega \le 2$. The exponentiation in Eq.~\eqref{overrelaxation} is 
obtained through the following representation for an element $u
\in$~SU(2):
\begin{equation}
\label{parametrization}
u = \cos(\frac{r}{2}) + i \left( \vec{\sigma} \cdot \hat{r} \right) 
\sin(\frac{r}{2}) \,,
\end{equation}
where $\hat{r}=\vec{r}/|\vec{r}|$.

\begin{figure}
\includegraphics[width=\textwidth]
{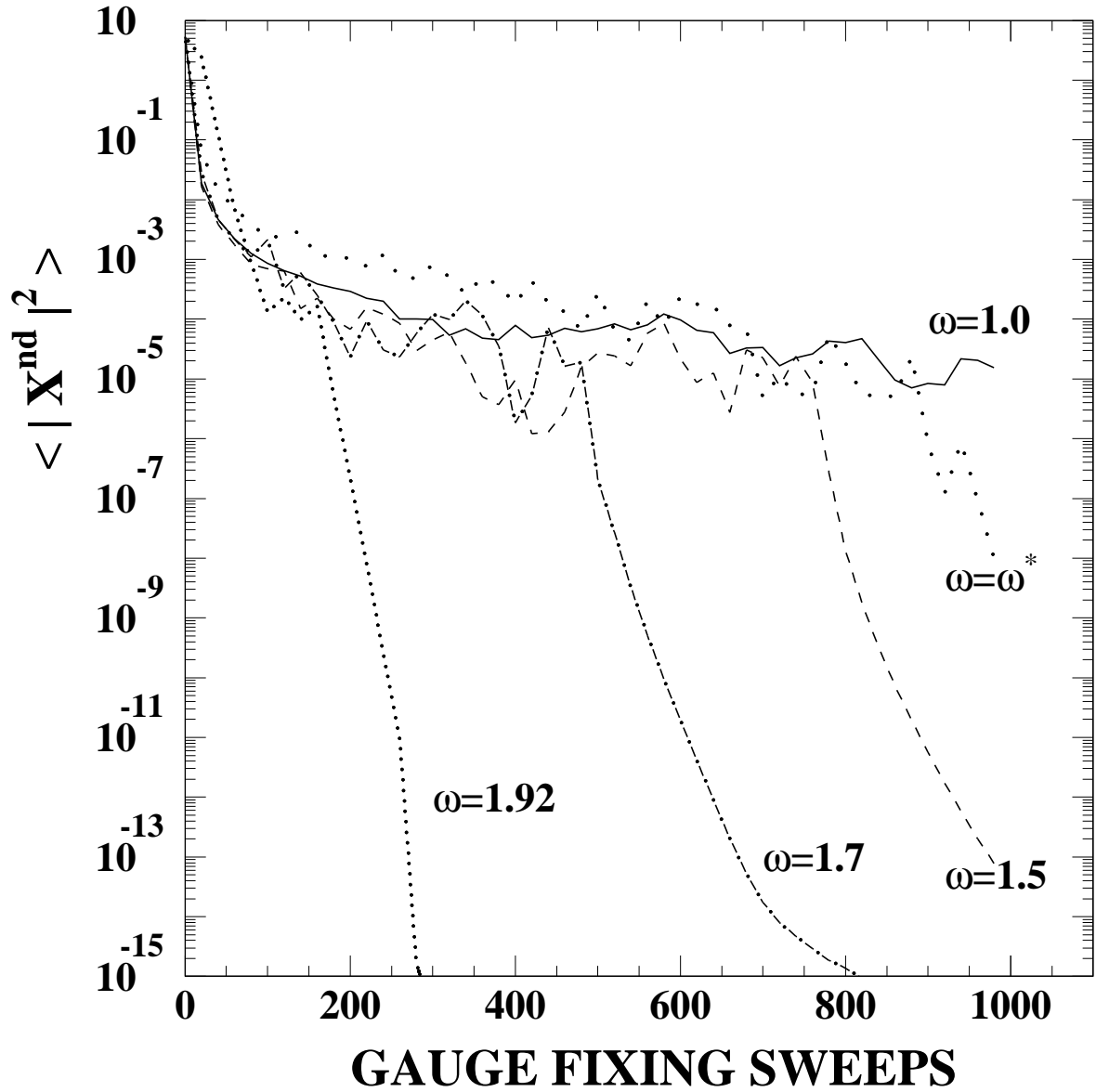}
\caption{Efficacy of gauge fixing defined by
Eq.~\protect\eqref{off-diag} as a function of the overrelaxation
parameter $\omega$ for the $L=16$ lattice. The case
$\omega=\omega^\ast$ corresponds to alternate $\omega=1.0$ 
with $\omega=2.0$ in the gauge fixing sweeps.}
\label{Fig:overrelaxation}
\end{figure}

In our Monte Carlo runs we used $\omega =1.7$.  However, we would like
to stress that it exists~\cite{Mandula90} an optimal overrelaxation
parameter $\omega_c$. Moreover for large lattice size $L$ it is
believed that
\begin{equation}
\label{omega_c}
\omega_c = \frac{2}{1+\frac{c}{L}}
\end{equation}
where the constant $c$ is problem dependent. As a matter of fact it
turns out that a better convergence can be obtained using values of
$\omega$ close to $1.9$ (see Figure~\ref{Fig:overrelaxation}). Indeed
we obtained $\omega_c \simeq 1.92$ for $L=16$. Inserting this value
into Eq.~\eqref{omega_c} we find $c \simeq 0.7$. It is remarkable that
our value for the constant $c$ agrees with the one relevant to the
Landau gauge fixing~\cite{Mandula90}.

%++++++++++++++++++++++++++++++++++++++++++++++++++++++++++++++++
%

%%%%%%%% FIGURES %%%%%%%%%%%%%%%%%
%
%Fig. 1
\begin{figure}
\begin{center}
\pspicture(0,0)(15,15)
%  linea   A-B
\psline[linewidth=2pt]{->}(   0.000,   0.000)(   4.000,   0.000)
\psline[linewidth=2pt](   0.000,   0.000)(   8.000,   0.000)
%  linea   B-C
\psline[linewidth=2pt]{->}(   8.000,   0.000)(  10.577,   3.059)
\psline[linewidth=2pt](   8.000,   0.000)(  13.154,   6.119)
%  linea   C-D
\psline[linewidth=2pt]{->}(  13.154,   6.119)(   9.154,   6.119)
\psline[linewidth=2pt](  13.154,   6.119)(   5.154,   6.119)
%  linea  D-E1
\psline[linewidth=2pt]{->}(   5.154,   6.119)(   3.968,   4.711)
\psline[linewidth=2pt](   5.154,   6.119)(   2.783,   3.304)
% linea   E1-FDOWN1
\psline[linewidth=2pt]{->}(   2.783,   3.304)(   4.783,   3.304)
\psline[linewidth=2pt](   2.783,   3.304)(   6.783,   3.304)
% linea   FDOWN1-FUP1
\psline[linewidth=2pt]{->}(   6.783,   3.304)(   6.783,   6.304)
\psline[linewidth=2pt](   6.783,   3.304)(   6.783,   9.304)
%  linea  FUP1-G
\psline[linewidth=2pt]{->}(   6.783,   9.304)(   7.195,   9.794)
\psline[linewidth=2pt](   6.783,   9.304)(   7.608,  10.283)
%  linea  G-H
\psline[linewidth=2pt]{->}(   7.608,  10.283)(   8.408,  10.283)
\psline[linewidth=2pt](   7.608,  10.283)(   9.208,  10.283)
%  linea  H-I
\psline[linewidth=2pt]{->}(   9.208,  10.283)(   8.692,   9.671)
\psline[linewidth=2pt](   9.208,  10.283)(   8.177,   9.059)
%  linea  I-FUP2
\psline[linewidth=2pt]{->}(   8.177,   9.059)(   7.377,   9.059)
\psline[linewidth=2pt](   8.177,   9.059)(   6.577,   9.059)
%  linea   FUP2-FDOWN2
\psline[linewidth=2pt]{->}(   6.577,   9.059)(   6.577,   6.059)
\psline[linewidth=2pt](   6.577,   9.059)(   6.577,   3.059)
% linea   FDOWN2-E2
\psline[linewidth=2pt]{->}(   6.577,   3.059)(   4.577,   3.059)
\psline[linewidth=2pt](   6.577,   3.059)(   2.577,   3.059)
%  linea  E2-A
\psline[linewidth=2pt]{->}(   2.577,   3.059)(   1.288,   1.530)
\psline[linewidth=2pt](   2.577,   3.059)(   0.000,   0.000)
%  label W
\uput{0}[0]{0}(  11.577,   3.059){{\Large$W$}}
%  label U
\uput{0}[0]{0}(   9.383,   9.304){{\Large$U_P$}}
%  label L
\uput{0}[0]{0}(   3.933,   2.295){{\Large$L$}}
\endpspicture
\end{center}
\vspace{2cm}
\caption{The connected correlator~\protect\eqref{rhoW} between the
plaquette $U_p$ and the Wilson loop. The subtraction appearing in the
definition of correlator is not explicitly drawn.}
\label{Fig:correlator}
\end{figure}
%
% Fig. 2
%
% Fig. 3
%
% Fig. 4
%
% Fig. 5
%
% Fig. 6
%
% Fig. 7
%
% Fig. 8
%
% Fig. 9
%
% Fig. 10
%
% Fig. 11
%
% Fig. 12
%
% Fig. 13
%
% Fig. 14
%
% Fig. 15
%
% Fig. 16
%
%Fig. 17

%%%%%%%% TABLES %%%%%%%%%%%%%%%%%
%
% Table 1
\begin{table}
\tabcolsep .2cm
\renewcommand{\arraystretch}{2}
\begin{center}
\begin{tabular}{|c||c|c|c|}
\hline
\multicolumn{4}{|c|}{{\bf fit parameters stability}} \\ \hline \hline
{$\mathrm{x_t^{min}}$} & {$\Phi$} & {$\mu$} & {$\chi^2/\mathrm{f}$} \\
\hline
 $1$ &    1.41345 ( 7599)  &  0.27123 ( 1246)  &  10.04140 \\ \hline
 $2$ &    1.29705 ( 6052)  &  0.34947 ( 2175)  &   1.17981 \\ \hline
 $3$ &    1.31656 ( 5568)  &  0.39057 ( 3695)  &   0.45213 \\ \hline
 $4$ &    1.35780 (11210)  &  0.41192 ( 6162)  &   0.44279 \\ \hline
 $5$ &    1.29926 (26320)  &  0.39328 (10050)  &   0.52273 \\ \hline
\end{tabular}
\end{center}
\caption{Fit parameters in Eq.~\protect\eqref{London} versus of the
number of discarded points in the $x_t$ direction.}
\label{Table:El_fit}
\end{table}
%
% Table 2
\begin{table}
\tabcolsep .2cm
\renewcommand{\arraystretch}{2}
\begin{center}
\begin{tabular}{|c||c|c|c|}
\hline
\multicolumn{4}{|c|}{{\bf fit parameters stability}} \\ \hline \hline
{$\mathrm{x_t^{min}}$} & {$\Phi_A$} & {$\mu_A$} &
{$\chi^2/\mathrm{f}$} \\
\hline
 $1$ &    1.56016 (21253)  &  0.20852 ( 2244)  &   4.35199  \\ \hline
 $2$ &    1.28796 (15337)  &  0.28366 ( 3880)  &   1.33579  \\ \hline
 $3$ &    1.24878 (10988)  &  0.35428 ( 6828)  &   0.41857  \\ \hline
 $4$ &    1.34201 (19876)  &  0.41971 (12542)  &   0.17380  \\ \hline
\end{tabular}
\end{center}
\caption{Fit parameters in Eq.~\protect\eqref{LondonAb} versus of the 
number of discarded points in the $x_t$ direction.}
\label{Table:ElAb_fit}
\end{table}


\begin{thebibliography}{99}
\bibitem{Nota4} This point is thoroughly discussed in the second paper
in Ref.~\protect{\cite{tHooft75}}.
\end{thebibliography}
\end{document}